\documentclass[structabstrac]{aa}

\usepackage{natbib}
\usepackage{graphicx}%
\usepackage{amssymb} 
\usepackage{amsmath}
\usepackage[usenames]{color}	
\usepackage{epstopdf}	
\usepackage{txfonts}

\DeclareGraphicsExtensions{.eps,.pdf,.ps}

\newcommand{\ds}{\displaystyle}

\newcommand{\teff}{T_{\rm eff}}
\newcommand{\teffsun}{T_{{\rm eff},\odot}}
\newcommand{\numax}{\nu_{\rm max}}
\newcommand{\numaxref}{\nu_{\rm ref}}

\newcommand{\taueff}{\tau_{\rm eff}}

\newcommand{\Ma}{{\cal M}_{\rm a}}
\newcommand{\MaO}{{\cal M}_{{\rm a},0}}

\newcommand{\deltanuref}{\Delta \nu_{\rm ref}}

\long\def\jumpover#1{{}}

\newcommand{\eq}[1] {Eq.~(\ref{#1})}

\newcommand{\diff}{{\mathrm{d}}}

\newcommand{\fig}[3]{
      \begin{figure}[ht]
        \centering
         \includegraphics[width=0.98\hsize]  {#1}
        \caption{#2}
        \label{#3}
        \end{figure} }

\newcommand{\eqn} [1] {
\begin{equation}#1
\end{equation}}
\newcommand{\eqna} [1] {
\begin{eqnarray}#1
\end{eqnarray}}

\usepackage[applemac]{inputenc}

\newcommand{\PI} {paper~I}

\begin{document}

\title{Stellar granulation as seen  in disk-integrated intensity}
\subtitle{II. Theoretical scaling relations compared with observations}
 \authorrunning{Samadi et al.}
  \titlerunning{Stellar granulation as seen  in disk-integrated intensity}

\date{\today}
\author{R. Samadi\inst{1}, K. Belkacem\inst{1}, H.-G. Ludwig\inst{2,3}, E. Caffau\inst{2,3}, T.L. Campante\inst{4}, G.R. Davies\inst{4},  T. Kallinger\inst{5,6}, M.N. Lund\inst{7}, B. Mosser\inst{1}, A. Baglin\inst{1}, S. Mathur\inst{8,9}, R.A. Garcia\inst{10}}

\institute{LESIA, Observatoire de Paris, CNRS UMR 8109, UPMC, Universit\'e Denis Diderot, 5 Place Jules Janssen, 92195 Meudon Cedex, France \and Zentrum f\"ur Astronomie der Universit\"at Heidelberg, Landessternwarte, K\"onigstuhl 12, D-69117 Heidelberg, Germany \and GEPI, Observatoire de Paris, CNRS UMR 8111, Universit\'e Denis Diderot, 5 Place Jules Janssen, 92195 Meudon Cedex, France  \and School of Physics and Astronomy, University of Birmingham, Edgbaston, Birmingham, B15 2TT, United Kingdom\and Institute for Astronomy (IfA), University of Vienna, T\"urkenschanzstrasse 17, 1180 Vienna, Austria \and Instituut voor Sterrenkunde, K.U. Leuven, Celestijnenlaan 200D, 3001 Leuven, Belgium \and Stellar Astrophysics Centre, Department of Physics and Astronomy, Aarhus University, Ny Munkegade 120, DK-8000 Aarhus C, Denmark \and High Altitude Observatory, NCAR, P.O. Box 3000, Boulder, CO 80307, USA \and  Space Science Institute, 4750 Walnut street Suite 205, Boulder, CO 80301, USA \and Laboratoire AIM, CEA/DSM CNRS  Universit\'e Paris Diderot IRFU/SAp, 91191 Gif-sur-Yvette Cedex, France}

\abstract
{A large set of stars observed by CoRoT and {{\it Kepler}} shows clear evidence for the presence of a stellar background, which is  interpreted to arise from surface convection, {{\it i.e.}}, granulation. These observations show that the characteristic time-scale ($\taueff$) and the root-mean-square (rms) brightness fluctuations ($\sigma$) associated with the granulation  scale  as a function of  the peak frequency ($\numax$) of the solar-like oscillations.  }
{We aim at providing a theoretical background to the observed scaling relations based on a model developed in the companion paper (hereafter \PI). }
{We computed for each 3D  model the theoretical power density spectrum (PDS) associated with the granulation as seen in disk-integrated intensity on the basis of the theoretical model published in \PI. For each PDS we derived the associated characteristic time ($\taueff$) and the rms brightness fluctuations ($\sigma$) and compared these theoretical values with the theoretical scaling relations  derived from the theoretical model and the measurements made on a large set of {{\it Kepler}} targets.  } 
{ 
 We derive theoretical scaling relations for  $\taueff$ and $\sigma$, which show the same dependence on $\numax$ as the observed  scaling relations. In addition, we  show that these quantities also scale as a function of 
  the turbulent Mach number ($\Ma$)  estimated at the photosphere. The theoretical scaling relations for $\taueff$ and $\sigma$ match   the observations well on a global scale. 
Quantitatively, the remaining discrepancies with the observations are found  to be much smaller than previous theoretical calculations made for red giants. 
 }
{Our modelling provides  additional theoretical support for the observed variations of $\sigma$ and $\taueff$ with $\numax$. It also highlights the important role of $\Ma$ in controlling the properties of the stellar granulation.   However, the observations made  with {{\it Kepler}} on a wide variety of stars cannot  confirm the dependence of our  scaling relations on $\Ma$.  Measurements of the granulation background and detections of solar-like oscillations in  a statistically sufficient number of cool dwarf stars will be required for confirming the dependence of the theoretical scaling relations with $\Ma$.}



\maketitle

\section{Introduction}

Since the launch of CoRoT (December 2006) and {\it Kepler} (March 2009), it is possible to accurately characterise  the properties of the stellar
granulation in other stars than the Sun \citep{Michel08,Ludwig09,Kallinger10b,Chaplin11,Mathur11}. A very large number of stars has been observed on a long term by CoRoT and {\it Kepler}. A large set of them clearly show both
solar-like oscillations and  a stellar background signal, which is
interpreted to be due to granulation at the surface of these stars.  These observations show that the characteristic time-scale of
the granulation scales  as the inverse of the peak frequency
($\nu_{\rm max}$) of the solar-like oscillation spectra detected in
these stars  \citep{Kallinger10b,Mathur11}. 
 In turn, the peak frequency $\nu_{\rm max}$ is known to scale as the cutoff-frequency $\nu_c$ of the atmosphere \citep{Brown91,Kjeldsen95,Stello09,Huber09,Mosser10,Kevin11}.
The observations of the stellar granulation background also reveal that the rms 
brightness fluctuations ($\sigma$)  associated with the the granulation
scales approximately as $\nu_{\rm max}^{-1/2}$  \citep{Mathur11,Chaplin11}.


The measurements made with {\it Kepler} on a large set of red giants have been compared in \citet{Mathur11} with theoretical calculations performed on the basis of the \citet{Ludwig06} {\it ab~initio} approach and using a grid of 3D hydrodynamical models of the surface layers of red giants computed with the STAGGER code \citep[see a detailed description for instance in][]{Trampedach04}. Although these theoretical calculations reproduce  the measured scaling relations rather well  in terms of their time-scale $\taueff$, and $\sigma$, large systematic differences were found with observations, however. While for the parameter $\taueff$ the dispersion to the scaling $\nu_{\rm max}^{-1}$ is very small,  a more appreciable dispersion is observed for  $\sigma$  with respect to the scaling $\nu_{\rm max}^{-1/2}$.  Eventually,  \citet{Mathur11} calculations were limited to red giant stars. 
For main-sequence (MS) stars, such a modelling has been performed and the results compared with observations for a limited set of stars only \citep{Trampedach98,Svensson05,Ludwig09,Guenther08}. 

 To justify the  scaling relation between $\taueff$  and $\nu_c$ (or equivalently $\numax$), \citet{Huber09} have conjectured that the granules move proportionally to the sound speed 
\citep[see also][]{Kjeldsen11}. 
The total brightness fluctuations $\sigma$ are known to scale as the inverse of the square root of the number of granules over half the stellar surface \citep[see e.g.][]{Ludwig06}.  
This number in turn scales as $\nu_c\, M/\teff^{3/2}$ \citep{Kjeldsen11,Mathur11}, where $\teff$ is the effective temperature and $M$ the mass of the star. However, $\sigma$ is expected to depend also on the intensity contrast of the granules and hence on their temperature contrast \citep[see e.g.][]{Ludwig09}.
 Therefore, the theoretical scaling relations proposed for $\sigma$ and $\taueff$ (the 'classical' scaling relations hereafter)  still partially rely on some simplified physical  assumptions and need to be completed.

\citet[\PI\, hereafter]{Samadi13a} proposed a theoretical model of the stellar granulation. This model predicts the power density spectrum associated with the relative variation of the disk-integrated  flux  due to the granulation at the surface of the star. 
Compared with the \citet{Ludwig06} {\it ab~initio} approach, this  theoretical model offers  the  advantage of testing separately several properties of the turbulent convection and can be extensively applied to a large set of stellar models.
Our aim here is to derive  theoretical scaling relations for $\taueff$ and $\sigma$ from this  model and to  compare them  with the observations made by {\it Kepler} on a large set of targets.  
For this purpose, we applied the theoretical  model of stellar granulation of \PI\, to a set of  3D hydrodynamical models of the surface layers of stars  with surface gravities ranging from $\log g= 1.5 $ to $\log g = 4.5$ and effective temperatures ranging from $\teff \simeq 6\,700$~K (F-type star) to $\teff= 4\,000$~K (K-type star). For each 3D model we computed the theoretical power density spectrum (PDS) associated with the granulation. From each spectrum we then extracted the characteristic time scale ($\taueff$) and the brightness fluctuations ($\sigma$) associated with the granulation in the same way as for the observations. We compare theses quantities with the  theoretical scaling relations derived  from  the theoretical model and  the characteristic times and  brightness fluctuations extracted from {\it Kepler} targets.
 


\section{Theoretical model}
\label{theoretical_model}

The theoretical model presented in \PI~ aims at modelling the power density spectrum (PDS) associated with the relative variations  of the bolometric flux emerging from the star in the direction of an observer and measured continuously  during a given duration.  Here we use the theoretical PDS   as a function of the frequency ($\nu$) given in \PI\,
\eqna{
{\cal F} (\nu) = \int_{0}^1 \diff \mu \, \int_{0}^{+\infty} \diff \tau\, e^{-2 \tau/\mu} \, \left ( \frac{\langle {B } \rangle_t  } {F_0} \right )^2 \;  {\cal F}_\tau (\tau,\nu) \, 
\label{F_nu_2}
}
with 
\eqna{
F_0  & = &   \int_{0}^1 \diff \mu \, \int_{0}^{+\infty} \diff \tau\, e^{- \tau/\mu} \,   \langle {B } \rangle_t (\tau)   \, \label{F0} \\
{\cal F}_\tau (\tau,\nu) & =  & { 2 \pi \, \tau_c \, \sigma_\tau^2   }   \, {S}_\Theta (\tau,\nu) \label{F_nu_tau}  \\
\sigma_\tau &  =  & \frac{12}{\sqrt{2}} \,  \sqrt { {\tau_g} \over {{\cal N}_g}  }\, \Theta_{\rm rms}^2 \label{sigma_tau} \\
\Theta_{\rm rms}& =  & {\Delta T_{\rm rms} \over T }\\
\tau_g  & = &  \kappa\,\rho\, \Lambda \label{tau_g} \\
{\cal N}_g   &=   &\frac{2 \pi R_s^2}{\Lambda^2} \label{N_g} \; ,
}
where $B$ is the Planck function, $\tau$ the mean optical depth, $\kappa$ the mean opacity, $\Delta T_{\rm rms}$ the rms of the temperature fluctuations, $T$ and $\rho$ the  stratification in temperature and density, respectively, $R_s$ the stellar radius, $\Lambda$ the granule characteristic size, $\tau_c$ the granule characteristic time, $\mu = \cos (\theta)$, $\theta$ the angle between the direction pointing toward the observer and the direction normal to the stellar surface, the symbol $\left< \right>_t$ stands for a time average,  and finally, ${S}_\Theta $ is a dimensionless ``source'' function whose expression is given in \PI.

 The term  ${\cal F}_\tau$ (\eq{F_nu_tau}) in the integrand of \eq{F_nu_2} stands for the PDS of the granulation as it would be seen at the optical depth $\tau$. 
In Eqs.~(\ref{F0})-(\ref{N_g}), $\tau_g$ corresponds to the  characteristic optical
thickness of the granules,  ${\cal N}_g $ to the average number
of granules distributed over  half of the photosphere (i.e. at $r = R_s$), and $\sigma_\tau$ to the rms brightnesses fluctuations
associated with the granulation spectrum as one would see  at the optical depth $\tau$.

The ``source'' function $S_\Theta$ requires a prescription for $\chi_k(\nu)$, the Fourier transform of the eddy-time correlation function (see \PI). The best fit with the solar granulation spectrum was obtained when an exponential form was adopted for $\chi_k(\nu)$. 

For comparison with the observations, we define $\sigma$ as the rms brightness fluctuations associated with the theoretical PDS (${\cal F} (\nu)$). The latter satisfies by definition the Parseval-Plancherel relation
\eqn{
\sigma ^2 =  \int_{-\infty}^{+\infty} \diff \nu \, {\cal F} (\nu) \, .
\label{sigma}
}
Following \citet{Mathur11}, we consider  $\taueff$ as the e-folding time  associated with the auto-correlation function (ACF) of the relative flux variations due to the granulation. 
Note that since the  ACF is also the  Fourier transform (FT) of the PDS, it is thus obtained by computing numerically the FT of ${\cal F} (\nu)$.

\section{Theoretical PDS across the HR diagram}
\label{PDS_HR}

\subsection{Grid of 3D hydrodynamical models}
\label{3Dmodels}
The 3D models used in this work are taken from the CIFIST grid \citep{Ludwig09b} and have been computed with the CO$^5$BOLD code \citep{Freytag12}. 
  The adopted chemical mixture is  similar to the solar chemical composition proposed by \citet{Asplund05}. 
Details about the CIFIST grid are given in \citet{Ludwig09b}. Of the 3D models of the CIFIST grid we considered only those with a solar metal abundance.  Their characteristics  are given in Table~\ref{tab:3Dmodels}. 

\begin{table*}
\caption{Characteristics of the 3D hydrodynamical models and associated parameters (see text). }
\center
\begin{tabular}{l|cccccc|cc|cc}
\hline
\hline
  Label  &     $\teff$   &   $\delta \teff$   &   $\log g$ &  $\Ma$ & $\Theta_{\rm rms}$ & $\numax$   &   $R_s$   &   $M$   &    $\sigma$   &   $\taueff$    \\
&  [K] & [K] & [cm/s$^2$] & & [\%] & [$\mu$Hz] & [$R_\odot$] & [$M_\odot$] & [ppm] & [s]   \\ \hline 
d3t40g15mm00n02 &   4018 &  24 & 1.50 & 0.323 &  4.75 &     4.30~~~~~  & 39.82 &  1.83 &     1.40~$10^{  3}$ &     9.06~$10^{  4}$\\
d3t45g25mm00n01 &   4476 &  10 & 2.50 & 0.274 &  3.72 &     4.07~$10^{  1}$ & 10.85 &  1.36 &     3.17~$10^{  2}$ &     1.09~$10^{  4}$\\
d3t45g40mm00n01 &   4477 &   8 & 4.00 & 0.178 &  2.13 &     1.29~$10^{  3}$ &  1.29 &  0.61 &     1.84~$10^{  1}$ &     5.11~$10^{  2}$\\
d3t48g32mm00n01 &   4775 &  12 & 3.20 & 0.260 &  3.45 &     1.98~$10^{  2}$ &  4.89 &  1.38 &     1.11~$10^{  2}$ &     2.49~$10^{  3}$\\
d3t50g20mm00n2 &   4552 &  16 & 2.00 & 0.320 &  4.62 &     1.28~$10^{  1}$ & 33.06 &  3.99 &     5.30~$10^{  2}$ &     2.91~$10^{  4}$\\
d3t50g25mm00n01 &   4969 &  18 & 2.50 & 0.378 &  4.66 &     3.86~$10^{  1}$ & 18.25 &  3.76 &     3.44~$10^{  2}$ &     1.01~$10^{  4}$\\
d3t50g30mm00n01 &   5037 &  18 & 3.00 & 0.301 &  4.14 &     1.21~$10^{  2}$ &  8.29 &  2.51 &     1.76~$10^{  2}$ &     3.53~$10^{  3}$\\
d3t50g35mm00n01 &   4924 &  13 & 3.50 & 0.245 &  3.17 &     3.88~$10^{  2}$ &  3.53 &  1.44 &     6.94~$10^{  1}$ &     1.32~$10^{  3}$\\
d3t50g40mm00n01 &   4955 &  11 & 4.00 & 0.207 &  2.44 &     1.22~$10^{  3}$ &  1.48 &  0.80 &     3.24~$10^{  1}$ &     4.88~$10^{  2}$\\
d3t50g45mm00n04 &   4981 &  12 & 4.50 & 0.172 &  1.98 &     3.86~$10^{  3}$ &  0.85 &  0.83 &     1.13~$10^{  1}$ &     1.77~$10^{  2}$\\
d3t55g35mm00n01 &   5432 &  22 & 3.50 & 0.308 &  4.06 &     3.69~$10^{  2}$ &  3.84 &  1.70 &     1.28~$10^{  2}$ &     1.20~$10^{  3}$\\
d3t55g40mm00n01 &   5476 &  13 & 4.00 & 0.257 &  3.30 &     1.16~$10^{  3}$ &  1.64 &  0.98 &     6.45~$10^{  1}$ &     4.47~$10^{  2}$\\
d3t55g45mm00n01 &   5488 &  14 & 4.50 & 0.216 &  2.58 &     3.68~$10^{  3}$ &  0.89 &  0.92 &     2.40~$10^{  1}$ &     1.64~$10^{  2}$\\
d3t59g35mm00n01 &   5885 &  16 & 3.50 & 0.381 &  4.39 &     3.55~$10^{  2}$ &  3.84 &  1.70 &     2.02~$10^{  2}$ &     1.11~$10^{  3}$\\
d3t59g40mm00n01 &   5927 &  13 & 4.00 & 0.311 &  4.07 &     1.12~$10^{  3}$ &  1.77 &  1.14 &     1.08~$10^{  2}$ &     4.11~$10^{  2}$\\
d3t59g45mm00n01 &   5861 &  25 & 4.50 & 0.257 &  3.35 &     3.56~$10^{  3}$ &  0.96 &  1.07 &     3.90~$10^{  1}$ &     1.53~$10^{  2}$\\
d3t63g35mm00n01 &   6140 &  25 & 3.50 & 0.445 &  4.46 &     3.48~$10^{  2}$ &  3.91 &  1.76 &     2.16~$10^{  2}$ &     9.56~$10^{  2}$\\
d3t63g40mm00n02 &   6227 &  16 & 4.00 & 0.362 &  4.38 &     1.09~$10^{  3}$ &  1.93 &  1.36 &     1.25~$10^{  2}$ &     3.77~$10^{  2}$\\
d3t63g45mm00n01 &   6233 &  15 & 4.50 & 0.297 &  4.04 &     3.45~$10^{  3}$ &  1.05 &  1.27 &     5.47~$10^{  1}$ &     1.44~$10^{  2}$\\
d3t65g40mm00n01 &   6486 &  20 & 4.00 & 0.416 &  4.46 &     1.07~$10^{  3}$ &  1.99 &  1.44 &     1.37~$10^{  2}$ &     3.54~$10^{  2}$\\
d3t65g45mm00n02 &   6458 &  14 & 4.50 & 0.325 &  4.36 &     3.39~$10^{  3}$ &  1.11 &  1.42 &     6.50~$10^{  1}$ &     1.37~$10^{  2}$\\
d3t68g43mm00n01 &   6725 &  17 & 4.25 & 0.406 &  4.35 &     1.87~$10^{  3}$ &  1.45 &  1.36 &     9.92~$10^{  1}$ &     2.00~$10^{  2}$\\
d3gt57g44n57 &   5783 &  18 & 4.44 & 0.272 &  3.21 &     3.11~$10^{  3}$ &  1.00 &  1.00 &     3.86~$10^{  1}$ &     1.72~$10^{  2}$\\
\hline
\end{tabular}
\label{tab:3Dmodels}
\end{table*}

For comparison with the observations, it is convenient to introduce the peak frequency of the solar-like oscillations ($\nu_{\rm max}$). This is shown to scale as the star's acoustic cut-off frequency, i.e.  as $g/\sqrt{\teff}$  \citep{Kjeldsen95,Stello09,Huber09,Mosser10,Kevin11,Mosser13}. Accordingly, we determined $\numax$ for each 3D model with the following scaling:
\eqn{
\numax = \numaxref \, { g \over g_\odot} \, \sqrt{ T_{\rm eff,\odot} \over \teff } \; , 
\label{numax}
}
where $\numaxref = 3\,106~\mu$Hz \citep[as in][]{Mosser13}, $\log g_\odot = 4.438$ (cm/s$^2$) and $T_{{\rm eff},\odot} = 5\,777$~K.
The values of  $\numax$ associated with each 3D model are given in Table~\ref{tab:3Dmodels}.
The positions of the 3D models in the plane $\teff - \numax$ are displayed in Fig.~\ref{HR}.

To calculate the theoretical PDS we needed  to know the radius of the star ($R_s$).   We determined for each 3D model the associated radius using a grid of standard stellar models computed with the CESAM2k code \citep{Morel08}. The stellar 1D models have the same chemical composition as the 3D models. The radius and mass associated with each 3D models are given in Table~\ref{tab:3Dmodels}.

\fig{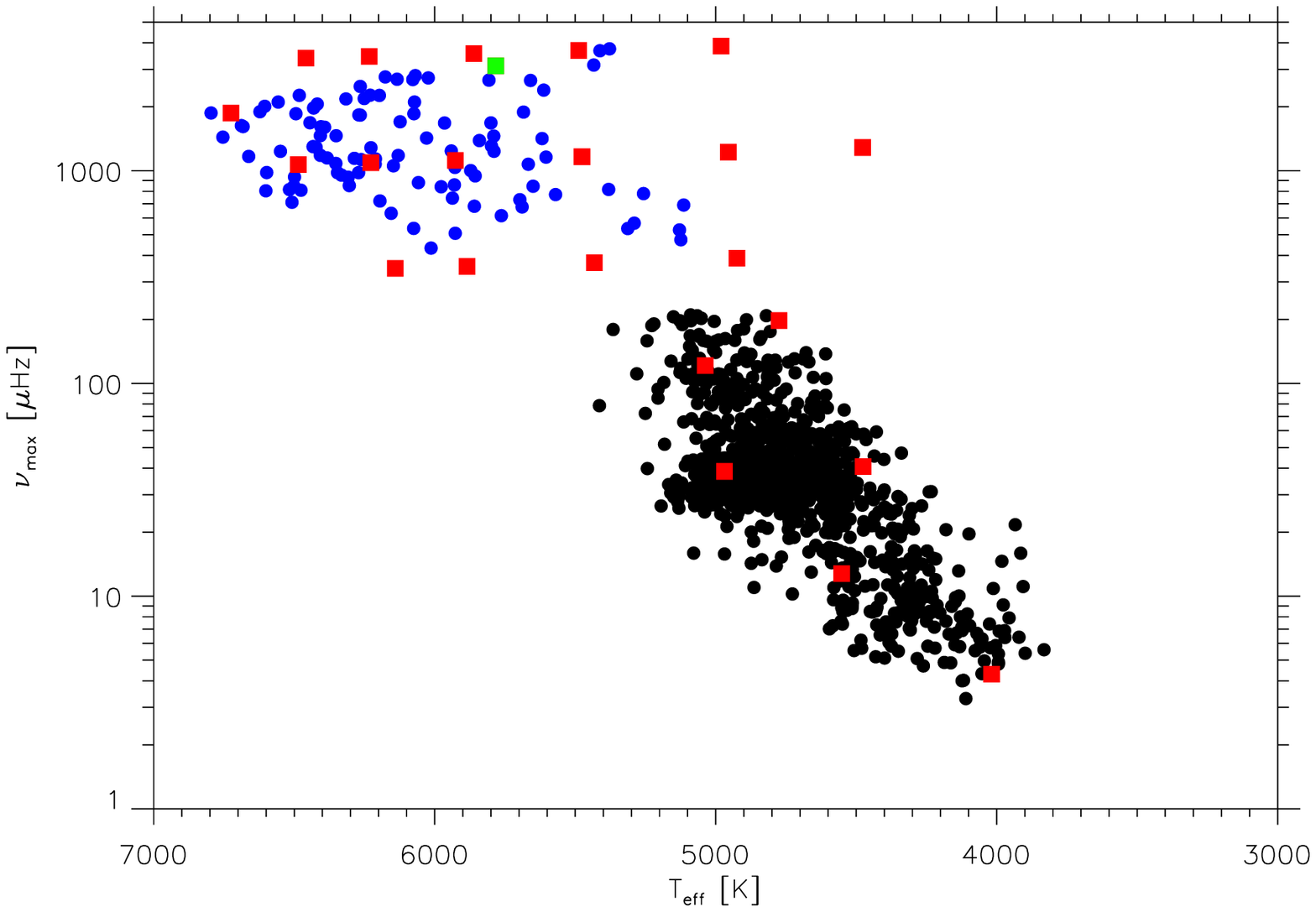}{ $\numax$  as a function of $\teff$. The filled red squares correspond to the location of the 3D hydrodynamical models in the  plan $\numax-\teff$ (see also Table~\ref{tab:3Dmodels}), the filled blue circles to the {\it Kepler} sub-giant and MS targets, and the black ones to the {\it Kepler} red giants (see Sect.~\ref{The observations}). The filled green square shows the position of our 3D solar model (the last model in Table~\ref{tab:3Dmodels}). }{HR}

\subsection{Calculations of the theoretical PDS}

 We computed the PDS of the granulation ($\cal F$) according to Eqs.~(\ref{F_nu_2})-(\ref{N_g}). The different quantities involved in the theoretical model  were obtained as detailed in \PI~from  3D hydrodynamical models of the surface layers of stars. 

The theoretical model involves three free parameters: $\beta$, $\lambda$, and $\zeta$.
The first controls the granule sizes, the second their characteristic time, and the last one  the characteristic wave-number, which separates the   inertial-convective range  from the inertial-conductive range  associated with the spectrum of the temperature fluctuations (see details in \PI). 
In \PI\, we calibrated ~the  three parameters using the observed solar granulation spectrum together with constraints from the resolved image of the solar granulation.  
The calibration gave $\lambda=0.30$, $\beta=3.42$ and $\zeta = 5$ (see \PI\, Sect.~4).  These values were used  for all calculations presented here.

For each 3D model, we computed the associated PDS and then derive the associated rms brightness fluctuation $\sigma$ (\eq{sigma}) and the characteristic time $\taueff$ as defined in Sect.~\ref{theoretical_model}. Theoretical values of $\taueff$ and $\sigma$ are given in Table~\ref{tab:3Dmodels} and are compared with the observations in Sect.~\ref{comparison}.


\section{Theoretical scaling relations}

It is observationally established  \citep{Kallinger10b,Mathur11,Chaplin11} that the
characteristic time $\taueff$ associated with granulation  varies
from one star to another approximately as the inverse of  the peak frequency of the solar-like
oscillations ($\nu_{\rm max}$). \citet{Brown91} and \citet{Kjeldsen95} have  conjectured that $\nu_{\rm max}$ scales as the star cut-off frequency, $\nu_c \propto g / \sqrt{T_{\rm eff}}$.  This relation was shown to work for a larger variety of stars \citep{Bedding03,Stello09,Huber09,Mosser10} and  its underlying physical origin has been explained recently by \citet{Kevin11}.
For the rms brightness fluctuation $\sigma$, the observations show that $\sigma$ roughly scales as $\left (\numax \right) ^{-1/2}$ \citep{Mathur11,Chaplin11}. 
As shown below, the theoretical model for the stellar granulation presented in Sect.~\ref{theoretical_model} predicts the same dependence of $\sigma$ and $\taueff$ on $\numax $. However, we will establish  that $\sigma$ and $\taueff$ are also expected to scale as a function of the Mach number.

 

The integrand of \eq{F_nu_2} is highest close to the
photosphere ({\it i.e.} around the optical depth $\tau = 2/3$, or equivalently around $T=T_{\rm eff}$). Therefore the rms brightness fluctuations $\sigma$ of the stellar granulation as well as its associated characteristic time $\taueff$ are closely controlled by  $\tau_c$  and $\sigma_\tau$  at the photosphere,  where  $\tau_c$  and  $\sigma_\tau$ are  the granule life-time and the rms brightness fluctuation at the  optical depth $\tau$ (see Sect.~\ref{theoretical_model}).

\subsection{Scaling for the characteristic time $\taueff$}
\label{scaling_taueff}

The characteristic time $\tau_c$ is by definition proportional to
$\Lambda/w_{\rm rms}$ (see \PI),  where $w_{\rm rms}$ is the rms of the vertical component of the velocity. Let $\Ma= w_{\rm rms}/c_s $ be the turbulent Mach number, where
  $c_s$ is the sound speed (both are evaluated at
the photosphere). By hypothesis, $\Lambda$  varies as
$H_p$, which scales as $T_{\rm eff}/g$. Furthermore,  $c_s$ varies at
the photosphere as $\sqrt{T_{\rm eff}}$. 
Accordingly, $\taueff$ is expected to scale as $\Ma \, \sqrt{T_{\rm eff}} / (g
\Ma) \propto (\Ma \, \nu_c)^{-1} $, and since $ \nu_c \propto \nu_{\rm
  max}$, we expect that
\eqn{
\taueff \propto { {1} \over {\Ma \, \nu_{\rm max} }} \; .
\label{scal_taueff}
}


\subsection{Scaling for the granulation amplitude  $\sigma$}
\label{scaling_sigma}

The expression for $\sigma_\tau$ (\eq{sigma_tau}) involves three characteristic
quantities:
\begin{itemize}
\item ${\cal N}_g$: the average number of granules over the stellar
  surface (\eq{N_g}); 
\item $\Theta_{\rm rms} = \Delta T_{\rm rms}/T$: the  rms of the (relative) temperature fluctuations;
\item $\tau_g$: the optical thickness of the granules (\eq{tau_g}).
\end{itemize}

From \eq{N_g} and the scaling relation for $\Lambda$ (see Sect.~\ref{scaling_taueff} above), one   easily obtains that   ${\cal N}_g$ scales as $\numax ~ M ~
\teff^{-3/2}$. 

\begin{figure}[ht]
\centering
 \includegraphics[width=0.98\hsize]  {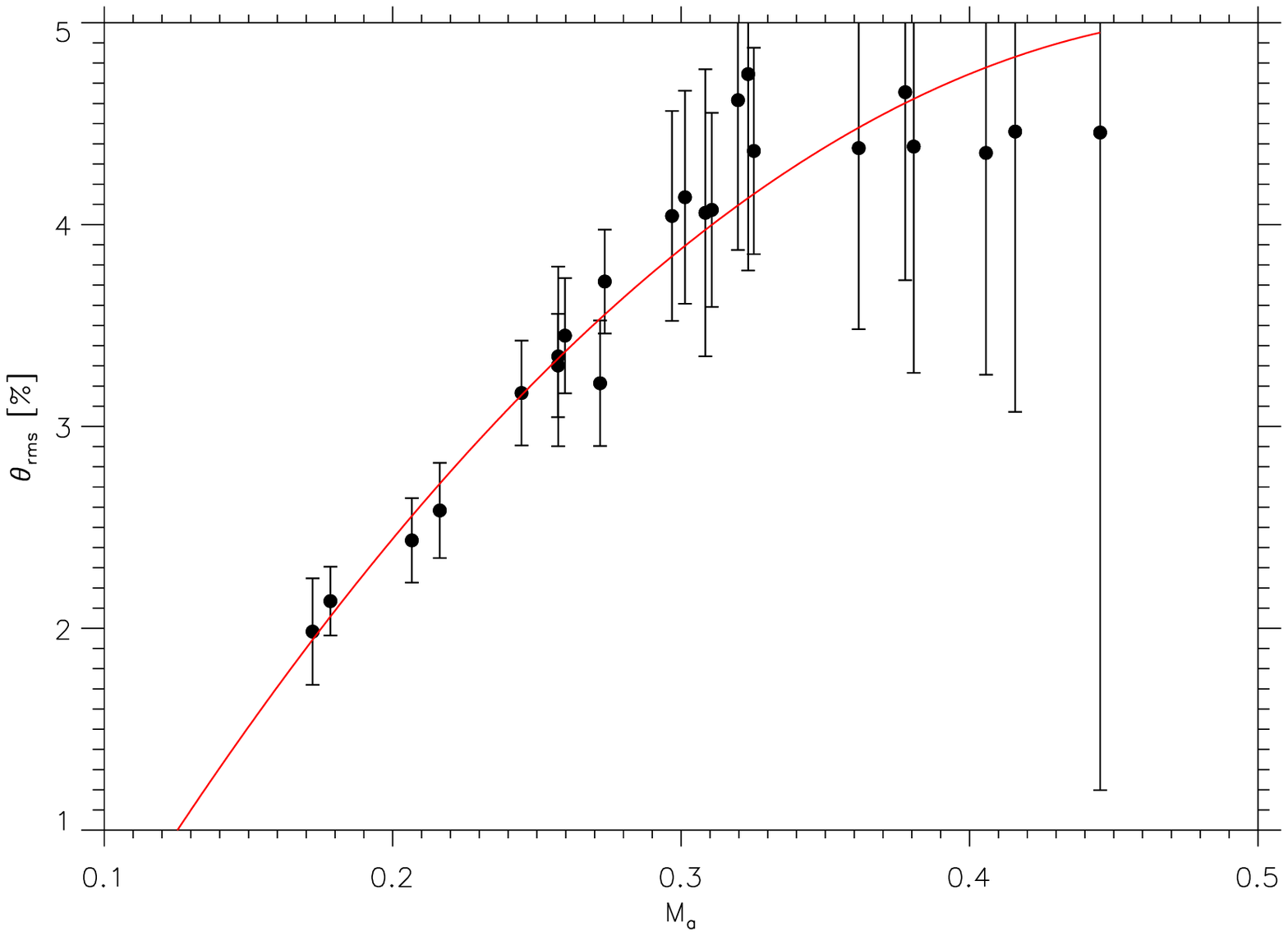}
 \includegraphics[width=0.98\hsize]  {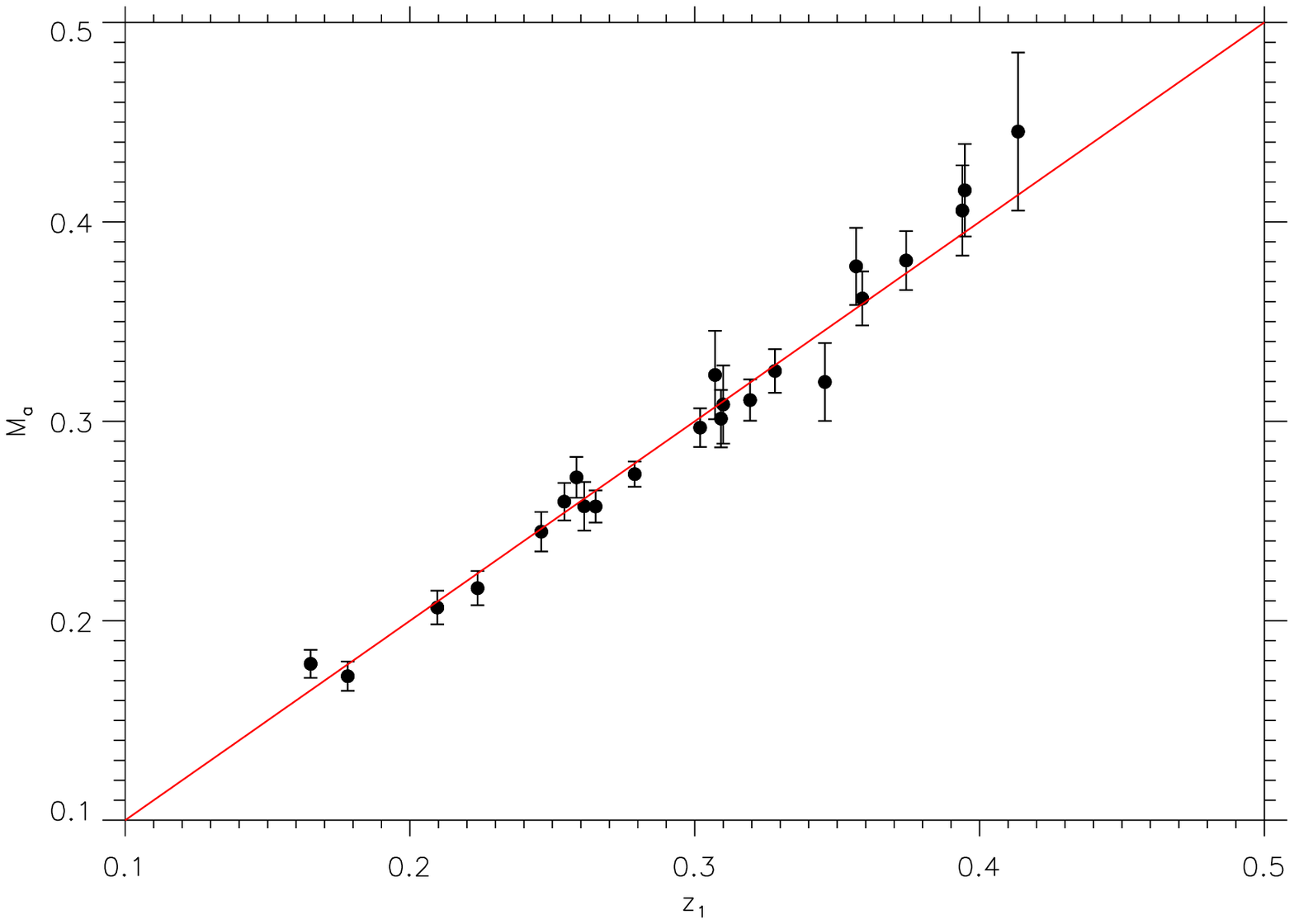}
\caption{{\bf Top:} Relative temperature fluctuation $\Theta_{\rm rms}$ (in \%) as a function of the Mach number $\Ma$. The filled circles  correspond to the values obtained for each 3D model (see Table~\ref{tab:3Dmodels}). Theses values were computed  at the photosphere (i.e. at the optical depth $\tau=$2/3). The red curve corresponds to the polynomial function given by \eq{theta_scaling}. The two upper points that deviate most from the polynomial function correspond to the two 3D models with  $\log g \leq 2$, i.e. the most evolved RG 3D models of our grid. {\bf Bottom:} $\Ma$ as a function of the quantity $z_1$ given by \eq{Ma3D}.  The red line corresponds to a linear scaling in $z_1$. 
}
\label{fig:theta:Ma}
\end{figure}

For $\Theta_{\rm rms}$  we can derive the following relation between $w_{\rm rms}$ and
the relative temperature fluctuations \citep[e.g.]{Cox68} following the
mixing-length theory:
\eqn{
w_{\rm rms}^2 =  g \, \Lambda \, \theta_{\rm rms}  \; .
\label{w_theta}
}
Since $\Lambda \propto T/g$ and $c_s \propto T^{1/2}$,  \eq{w_theta} yields  that  $\Theta_{\rm rms}$  scales as   $\Ma^2$. 
However, using a grid of stellar 3D models (see Sect.~\ref{3Dmodels}),  we found that  $\Theta_{\rm rms}$  varies with $\Ma$ in a more complicate manner. This is illustrated in  Fig.~\ref{fig:theta:Ma} (top panel), where $\Theta_{\rm rms}$ is plotted as a function of $\Ma$. 
Except for  the two red-giant 3D models with $\log g \leq 2$ (see Table~\ref{tab:3Dmodels}),   $\Theta_{\rm rms} (\Ma)$  can   be fitted very well by a second-order polynomial function
\eqn{
{\Theta_{\rm rms} \over \Theta_0 }  \equiv   f(\Ma) =  a_0 + a_1 \, \left ( { \Ma \over \MaO} \right ) + a_2 \, \left ( { \Ma \over \MaO} \right ) ^2  \; , 
\label{theta_scaling}
}
where the coefficients ($a_0$, $a_1$, $a_2$) are obtained by least-squares adjustment, and where we have  defined the coefficients $\Theta_0$=3.2~\% and $\MaO$= 0.26. 
The value  $\Theta_0$ corresponds  to the value associated with the solar 3D model (the last model listed in Table~\ref{tab:3Dmodels}), while the value  $\MaO$= 0.26 is the one obtained by adjustment in Sect.~\ref{scaling_Ma}. 
 The fit yields $a_0=-0.67 \pm 0.21 $, $a_1= 2.30 \pm 0.43$ and $a_2= -0.59 \pm 0.22 $  and is shown in Fig.~\ref{fig:theta:Ma} (top panel).   It is interesting to note that \citet{Tremblay13} have recently found a qualitatively similar dependence between the relative intensity contrast of the granules and $\Ma$.

Finally, the optical thickness of the granule $\tau_g$ is expected to vary slowly from one star to another. Indeed, we aim to evaluate this
quantity near the photosphere, i.e. close to the unit optical depth. The optical thickness of the granule is by definition
\eqn{
\tau_g = \Lambda \kappa \rho = \beta \, H_p \,
\kappa \rho \; , 
}
where $\beta$ is the free parameter introduced in \PI.
The optical depth is given by  $\tau (r) = \ds \int_r^{+\infty} {\rm d}r \, 
\kappa \, \rho$.
At the photosphere,  the pressure scale-height, $H_p$, is of the same order as the
density scale height $H_\rho$. If we assume that the opacity and the
density scale height vary slowly in the atmosphere, we obtain that at
a given optical depth  $\tau \propto \kappa \, \rho \, H_p \propto \tau_g$. 
The photosphere corresponds by definition to the optical depth $\tau=
2/3$. Therefore, if we evaluate $\tau_g$ at the photosphere, we then
expect that $ \tau_g$ remains constant from a star to another. 
 
When we combine the scaling relations found for ${\cal N}_g$ and $\Theta_{\rm rms}$, we find
the following scaling:
\eqn{
\sigma  \propto { f^2(\Ma)  \over \nu_{\rm max}^{1/2}}  \, 
{ \teff^{3/4} \over  {M^{1/2}}} \; ,
\label{sigma_scaling}
}
where the function $f(\Ma)$ is given by \eq{theta_scaling}. 
 The term $ \teff^{3/4} \,  {M^{-1/2}} \, \nu_{\rm max}^{-1/2}$ in \eq{sigma_scaling} corresponds to the classical scaling relation \citep{Kjeldsen11,Mathur11}, and comes basically from the fact that $\sigma$ scales as the inverse of the square root of the number of granules over  the stellar surface \citep[see e.g.][]{Ludwig06}.
Our theoretical model then consistently results in the same dependence as the  classical scaling relation. However, we find here that  $\sigma$ also scales with a function of the Mach number.  
It is worth noting that  for MS stars, the term $(\teff^{3/4} / M^{1/2})$ varies slowly. Indeed,
according to \cite{Noyes84}, $M \propto T^{1.8}$ for MS stars.

\subsection{Scaling for the Mach number $\Ma$}
\label{scaling_Ma}

To compare the scaling relations established for $\taueff$ (\eq{scal_taueff}) and $\sigma$  (\eq{sigma_scaling}) with the observations, it is  necessary to derive the Mach number $\Ma$ as a function of some fundamental parameters of the star.  As a guideline, we first derived such a scaling on the basis of simple physical assumptions. We  then derived a more appropriate scaling using the grid of 3D models.

We first established a scaling for the  flux of kinetic energy $F_{\rm kin} \approx \rho \, w_{\rm rms}^3$. 
In the framework of the mixing-length approach, it can be shown that $F_{\rm kin}$ is roughly proportional to the convective flux $F_c$. In the layer where the granulation is observed, the total energy flux, $F_{\rm tot}$,  is no longer  transported dominantly by convection. However,  to derive an expression for $F_{\rm kin}$  that depends only on the
surface parameters of the star, we assumed that  the entire energy is transported by
convection;  that is  $ F_c \approx F_{\rm tot} = \sigma \, T_{\rm eff}^4  $, where $\sigma$ is the Stefan-Boltzmann constant.  Accordingly, at the photosphere we have  $\Ma \equiv w_{\rm rms}/ c_s \propto \teff^{5/6} \, \rho_s^{-1/3}$ where $\rho_s$ is the density at the photosphere. 

We now need to  establish a scaling for $\rho_s$. As done in Sect.~\ref{scaling_sigma}, we approximated the optical depth as $\tau \approx \kappa \, \rho \, H_p$.
Since our goal is to derive the scaling relation for $\Ma$ at the photosphere, which by definition is such that  $\tau = 2/3$, we then obtain that $\rho_s$ must  scale as $ g \, \teff^{-1}\,  \kappa^{-1} $.

For the low-mass stars  we are interested in, which have typical temperatures ranging between 4\,000 and 6\,000 $K$, the dominant opacity source is ${\rm H}^-$. 
Using tables, it has been possible to show that the related opacity follow the power law \citep[e.g.][]{Hansen94}
\begin{equation}
\kappa \propto \rho^{1/2} \, T^9 \, .
\end{equation} 

We finally establish 
\eqn{
\Ma \propto z_0 \equiv \left (\teff \over \teffsun \right ) ^3 \, \left (g  \over g_\odot \right ) ^{-2/9} \; . 
\label{Ma}
}
Guided by this scaling, we fitted  the following analytical expression  on our set of values of $\Ma$: 
\eqn{
\Ma = z_1 \equiv  \left (\teff \over \teffsun \right ) ^{a} \, \left (g  \over g_\odot \right ) ^{-b} \, \MaO \; ,
\label{Ma3D}
}
where $a$, $b$ and $\MaO$  are  coefficients obtained by least-squares adjustment.   The best fit is obtained with $ a= 2.35 \pm 0.09$, $b= 0.152 \pm 0.007 $ and $\MaO=0.258 \pm 0.003 $.   
The overall satisfactory agreement of the scaling given by \eq{Ma3D} with the individual values for $\Ma$ (Table \ref{tab:3Dmodels}) is illustrated in Fig.~\ref{fig:theta:Ma} (bottom panel).

\subsection{Comparison with individual theoretical values}
\label{comparision_individual}

We compared the theoretical scaling relations derived for $\taueff$ (Sect.~\ref{scaling_taueff}) and $\sigma$ (Sect.~\ref{scaling_sigma}) with the individuals values obtained with  our grid of 3D models (see Sect.~\ref{PDS_HR} and Table \ref{tab:3Dmodels}) as well as with the classical scaling relations. 

\begin{figure}[ht]
\centering
 \includegraphics[width=0.98\hsize] {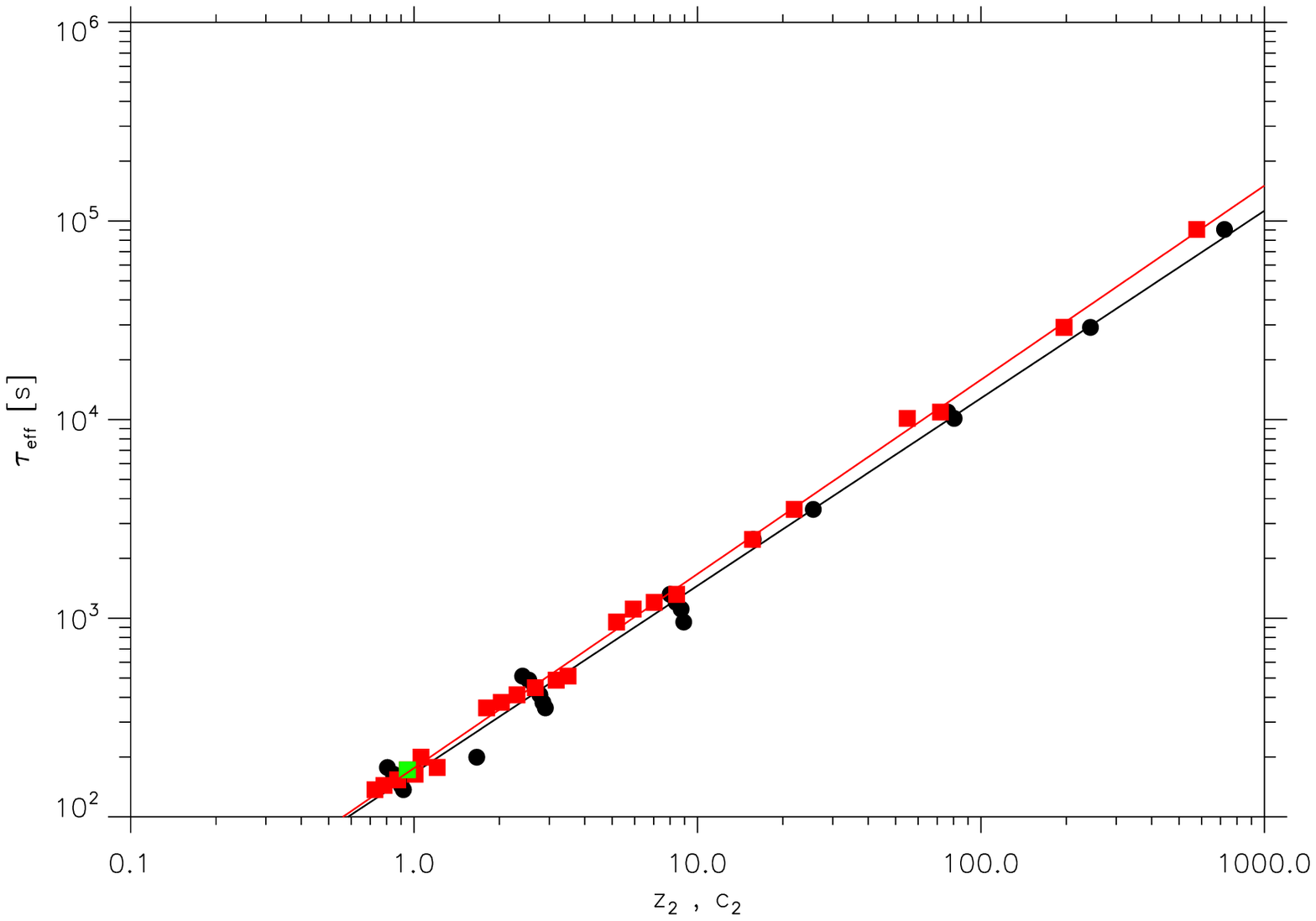}
\includegraphics[width=0.98\hsize] {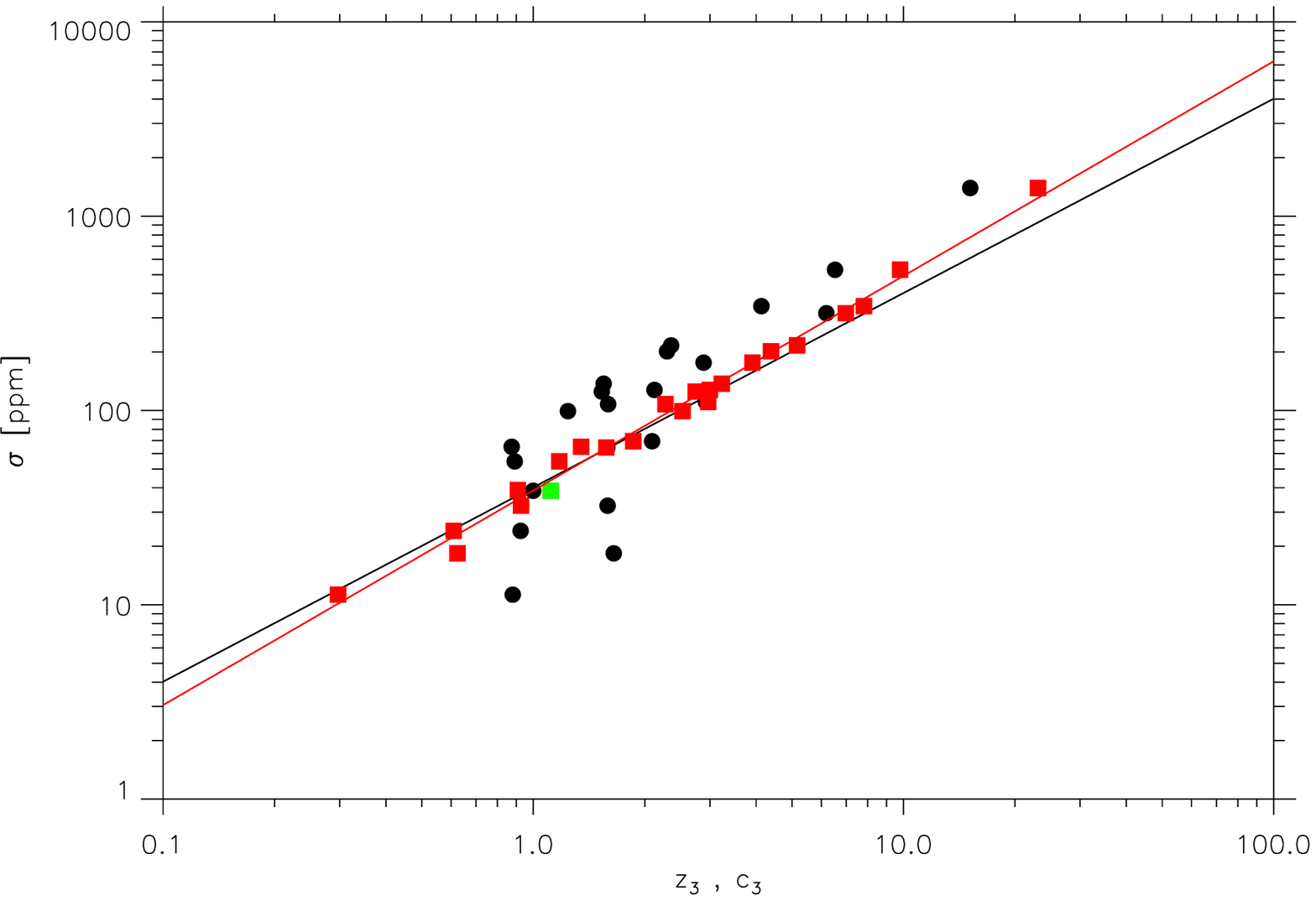}

\caption{ {\bf Top:} Theoretical values of $\taueff$ as a function of the  quantity $z_2$ (red squares) given by \eq{scal_taueff2} and as a function of the classical scaling $c_2 \equiv (\numaxref/\numax)  $ (filled black circles). The symbols correspond to the individual theoretical values obtained with our grid of 3D models.  The red curve corresponds  to a power law of the form  $ z_2^{p}$ where the slope $p = 0.98$ is obtained by fitting the red squares, while the black line is a power law of the form  $ c_2^{n}$ where the slope $n = 0.94$ is obtained by fitting the black circles. {\bf Bottom:}   Theoretical values $\sigma$ as a function the quantity $z_3$ given by \eq{sigma_scaling_2} and as a function of the classical scaling $c_3$ (\eq{c_3}). The symbols correspond to  the individual theoretical values obtained with our grid of  3D models.  The red curve is a power law of the form  $ z_3^{p}$ where the slope  $p = 1.10$ is obtained by fitting the red squares, while the black line  is a linear scaling in $c_3$.  }
\label{fig:scaling_individual_values} 
 \end{figure}

We started with the scaling relations for $\taueff$. We recast the scaling relation given by \eq{scal_taueff} as
\eqn{
\taueff \propto z_2 \equiv \left (  \numaxref \over \numax \right ) \, \left (  \MaO  \over \Ma\right ) \; .
\label{scal_taueff2}
}
We have plotted in Fig.~\ref{fig:scaling_individual_values} (top panel) individual theoretical values  of $\taueff$ (red squares) as a function of the quantity $z_2$. 
The individual theoretical values of $\taueff$  (Tab.~\ref{tab:3Dmodels}) were found to scale as $z_2^p$ with $p$=0.98. 
This scaling is then close from the theoretical scaling relation $z_2 \propto (\numax \, \Ma )^{-1}$.  However, there is a slow deviation from a linear scaling with $z_2$, which must very likely be attributed to the various simplifications adopted in Sect.~\ref{scaling_taueff} to derive this scaling relation.
For comparison with the classical scaling relation for $\taueff$, we defined the quantity $c_2 \equiv (\numaxref/\numax) $. 
We have plotted in Fig.~\ref{fig:scaling_individual_values} (top panel) individual theoretical values  of $\taueff$ (black circles) as a function of $c_2$.
Theoretical $\taueff$ are found to scale as  $c_2^{n}$ with $n$= 0.94. The deviation from a linear scaling were therefore higher than with the scaling with $z_2$. Furthermore, for MS and sub-giant stars a much higher dispersion is obtained than with the new scaling relation.  This shows that for these stars the Mach number significantly influences   the characteristic time $\taueff$. On the other hand, for evolved stars, the variation of $\taueff$ is dominated by the variation of $\numax$ along the evolution, which is  basically related to rapid variation of the surface gravity (or equivalently the luminosity).

We turn now to the scaling relations for $\sigma$. The theoretical scaling relation given by \eq{sigma_scaling} is recast as
\eqna{
\sigma  \propto & z_3 \equiv  & \left ( {\teff \over \teffsun} \right ) ^{3/4} \, \left ( { M_\odot \over M} \right )^{1/2} \,   \left ( {\numaxref \over \numax}\right )  ^{1/2} \, 
\left ( { {f(\Ma)} \over {f(\MaO) } } \right )^2 \; .
\label{sigma_scaling_2}
}
As seen in Fig.~\ref{fig:scaling_individual_values} (bottom panel), individual theoretical values of $\sigma$ (red squares) vary  globally  according to the theoretical scaling relation given by \eq{sigma_scaling_2}. The scaling of $\sigma$ with $z_3$ is not fully linear since a fit gives $\sigma \propto z_3^{p}$ with $p=1.10$ (see Fig.~\ref{fig:scaling_individual_values}, bottom panel).
 This deviation from   a linear scaling with $z_3$ is shown in Appendix~\ref{degeneracy}  to arise for a large part from the considerable degeneracy between $R$ and $M$ that occurs for RG stars. For comparison with the classical scaling relation, we defined the quantity $c_3$ as
\eqn{ 
c_3  \equiv \left ( {\teff \over \teffsun} \right ) ^{3/4} \, \left ( { M_\odot \over M} \right )^{1/2} \,   \left ( {\numaxref \over \numax}\right )  ^{1/2} \; .
\label{c_3}
}
Theoretical values of $\sigma$ are plotted in Fig.~\ref{fig:scaling_individual_values} (bottom panel) as a function of $c_3$. Again, for the MS and sub-giant stars, a considerable dispersion is obtained w.r.t. the classical scaling relation $c_3$. The dispersion is substantially reduced when $\sigma$ is plotted as a function of the new scaling relation ($z_3$).

As a conclusion, despite the simplifications adopted for deriving the new scaling relations, they are found to match  the values of $\sigma$ and $\taueff$ derived from the theoretical PDS reasonably well. 
On the other hand, the classical scaling relations significantly departs from the theoretical values.

\section{Observations}
\label{The observations}

 Characteristics of stellar granulation in terms of time-scale ($\taueff$) and rms brightness fluctuations ($\sigma$) are typically extracted from the power spectrum of the intensity using various background models. 
Different analysis methods  are found in the literature. They mainly differ from each other in 1) the number of components  fitted in addition to the granulation component (e.g. activity, super-granulation, modes, misidentified components), and 2) the functional forms adopted for each component.

 A major source of uncertainty related to those methods arises from the kink that is more or less visible on the stellar background. Such a feature has been first identified in  the power spectrum of the solar irradiance data from the SOHO/VIRGO instrument at around 1~mHz \citep{Andersen98,Vazquez-Ramio05}. Similar features seem to be observed by {\it Kepler} on RG stars \citep{Mathur11}.  
 Furthermore, \citet{Karoff13}   recently analysed three MS stars for which this kink is visible and statistically significant. The physical origins of this kink are subjects of debates. They are  either attributed to the occurrence of bright points \citep[e.g.][]{Harvey93,Aigrain04},  the changing properties of the granules \citep{Andersen98}, a  second granulation population \citep{Vazquez-Ramio05}, or to faculae \citep{Karoff11}.

%
Because the origin of this second component is not yet clear, and  it is furthermore missing in the current theoretical models,  we considered  the high-frequency component to be part of the granulation spectrum and compared the characteristics of the whole granulation spectrum with our theoretical predictions.

To compare measured values of $\sigma$ and $\taueff$ with theoretical calculations,  we fitted the observed stellar backgrounds with the same functional form for all types of stars (MS stars to red giants).
 Of the different forms studied in \citet{Mathur11}, the   Lorentzian  function \citep[also named in this context the Harvey model,][]{Harvey85} results in values for $\taueff$ and the height of the granulation spectra that are enclosed by the different other methods of analysis investigated by \citet{Mathur11}. 
Accordingly,    we adopted   a Lorentzian  function  as in \citet{Mathur11} as a reference  for our comparisons with the theoretical calculations, 
 \eqn{
{ \cal P} (\nu) = {{H_g  } \over {1 + (2 \pi \, \taueff \, \nu)^2} } \;,
\label{lorenztian}
}
where the height $H_g$ and $\taueff$ were obtained as explained below. 
Note that prior to the calculation and the fit of the PDS, all light-curves were corrected following the procedures described in \citet{Garcia11}. 

For red giants, $H_g$ and $\taueff$ were determined with the method named 'COR' in \citet{Mathur11}  that was also used in \cite{Mosser12}.
Locally around $\numax$ in a frequency range equal to [0.15- 6]$\times \numax$, one Lorentzian component is enough for fitting the background. The scaling relation $\taueff \propto \numax^{-1}$ and the determination of the background $B_{\mathrm{max}}$ at $\numax$ provided  guess values of $\taueff$ and $H_g$. The seismic excess energy was modelled with a Gaussian with a FWHM and an amplitude that are also governed by scaling relations \citep[e.g.][]{Mosser10,Hekker11b}. All parameters were then iteratively determined.  The sample of red giants from which $H_g$ and $\taueff$ were extracted was the same as in \citet{Mosser12} or in \citet{Mathur11}.

More than about 500 of  the sub-giants and MS stars, observed during Kepler's survey phase ({\it i.e.}, based on one month of observations) were seen to be oscillating by {\it Kepler}.  Here, however, we have analysed a cohort of the global sample that have been selected for long-term follow-ups and have thus been observed for at least three months from Quarter 5 onwards. This cohort initially contained 196 stars. In our final cut, we have retained the results from 141 stars. 
Taking $\numax \approx$ 800 $\mu$Hz as the threshold separating sub-giant from MS stars, 108 targets of the 141 targets of the  sample  can thus be considered as  MS stars. 
Our analysis of their stellar backgrounds assumed a single Harvey-like profile (cf. \eq{lorenztian}) describing granulation, a flat component describing shot noise, and a Gaussian envelope describing the p-mode hump. The power spectra were fitted in a range starting at 100 $\mu$Hz and extending to the Nyquist frequency of {\it Kepler} short-cadence data (8.5~mHz). A maximum-likelihood approach was employed to determine $H_g$ and $\taueff$.   More details about this analysis can be found in  \citet{Karoff13}. 

The rms brightness fluctuations $\sigma$, associated with the form  given by \eq{lorenztian} was obtained according to
the relation $ \sigma^2 =   C_{\rm bol}^2 \, H_g / \taueff /4 $, where $C_{\rm bol}$ is a bolometric correction that scales  for the ${\it Kepler}$ bandpass as  $C_{\rm bol} = \left ({\teff /  T_0} \right) ^\alpha$, where $T_0 = 5934~$K and $\alpha = 0.8$  \citep[][see also \citet{Michel09}]{Ballot11b}. 

Measured values of $\taueff$ and $\sigma$ are compared in the next section with those obtained from the theoretical PDS, as explained in Sect.~\ref{theoretical_model}. However, in this comparison one must keep in mind that our measured values of $\taueff$ and $\sigma$ depend on the way the stellar background is modelled, in particular, whether or not the high-frequency component is included in the background model. As mentioned above,  relative uncertainties of  about  30\,\% (peak-to-peak) remain.

\section{Comparison with the observations}
\label{comparison}

\subsection{Characteristic time-scale, $\taueff$}
\label{comparision_taueff}



\begin{figure}[ht]
        \centering
         \includegraphics[width=0.98\hsize]  {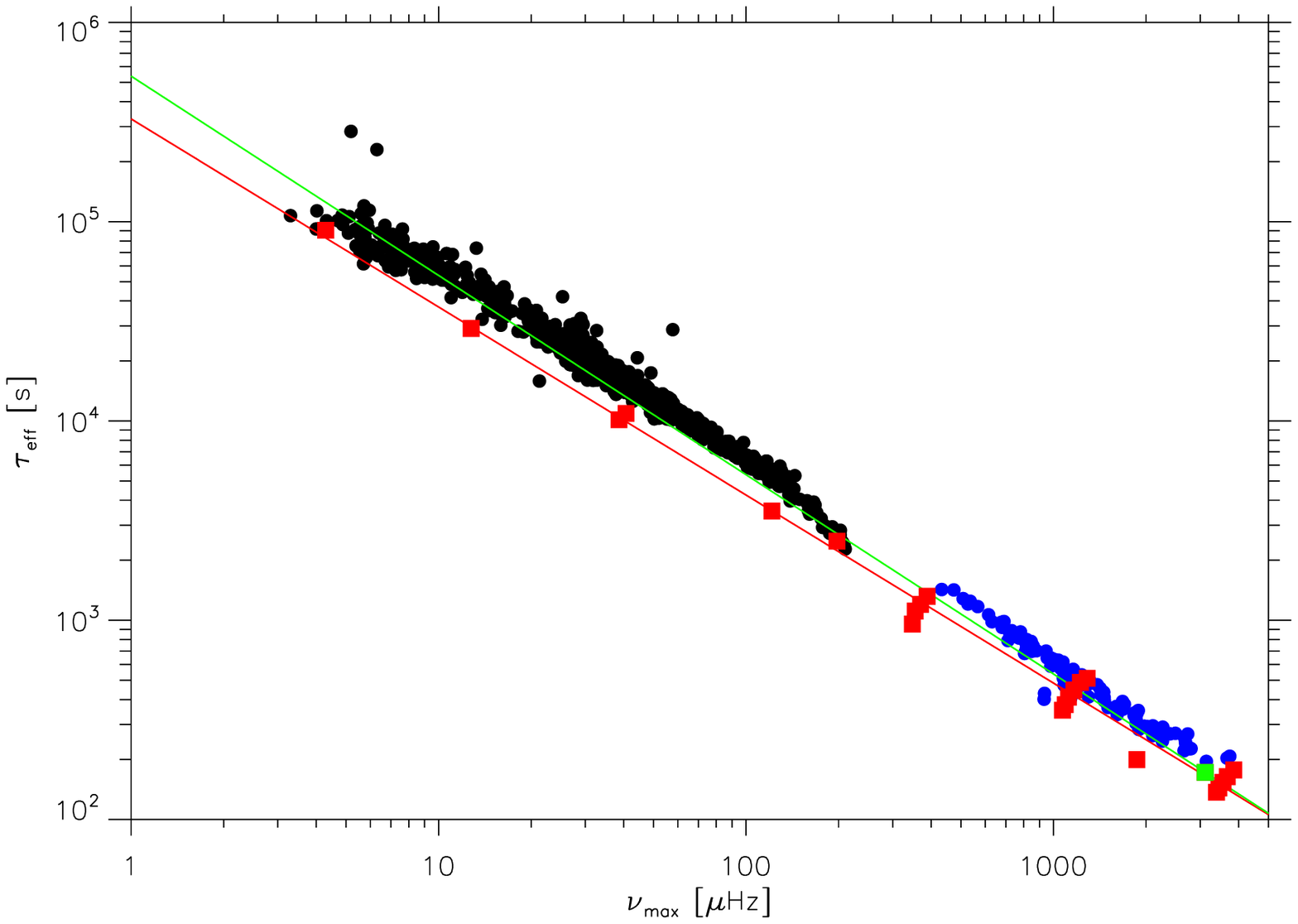}
         \includegraphics[width=0.98\hsize]  {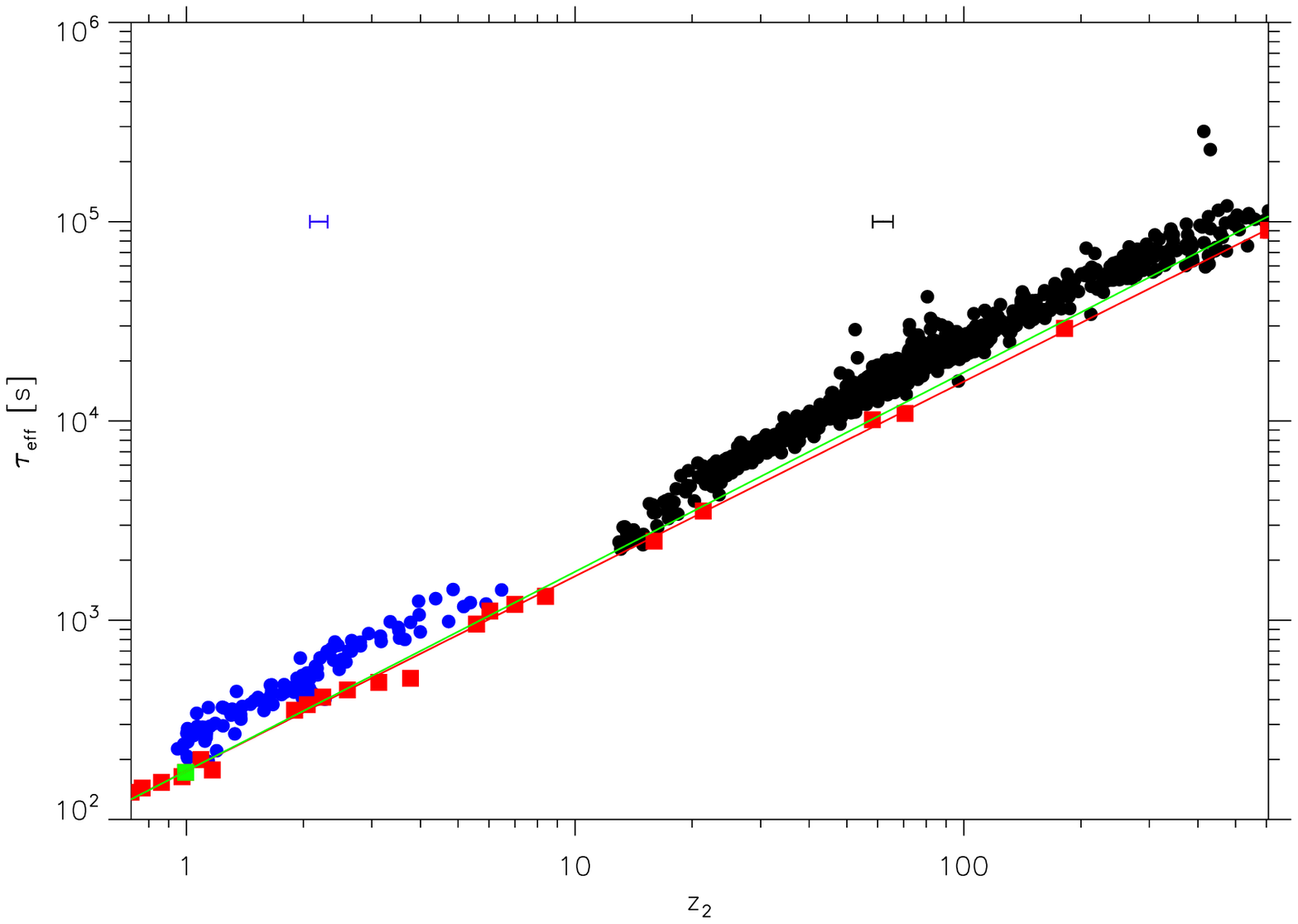}

\caption{{\bf Top:} Characteristic time $\taueff$ as a function of $\numax$. The dots have the same meaning as in Fig.~\ref{HR}.  The green curve correspond to a linear scaling in $\numax^{-1}$ while the red curve is a power law of the form $\numax^{p}$ with the slope $p=-0.94$ obtained by fitting the power law to the theoretical values of $\taueff$ (filled red squares). The filled green square corresponds to the value $\taueff = 173$~s found for the solar 3D model (the last model in Table~\ref{tab:3Dmodels}). {\bf Bottom:} $\taueff$ as a function of the scaling $ z_2 = ( \numax /\numaxref)   \, (\Ma/ \MaO) $ where  $\Ma$ is supposed to scale according to \eq{Ma3D}. The symbols have the same meaning as in Fig.~\ref{HR}.  The green line corresponds to a linear scaling in $z_2$, while the red curve  to a power law of the form  $ z_2^{p}$ where the slope $p = 0.98 $ is obtained by fitting the theoretical values of $\taueff$ (red squares).  The horizontal error bars show the uncertainty in $z_2$ associated with a typical uncertainty of 100~K (rms) in $\teff$. The blue horizontal error bar corresponds to a typical MS and the black one to a typical RG star.
} 
\label{fig:taueff_numax_Mascaling}

 \end{figure}

First, we compare in Fig.~\ref{fig:taueff_numax_Mascaling} (top panel) individual  theoretical values of $\taueff$ (red squares) with measured ones. Sub-giants and MS stars overlap  the theoretical and measured values of $\taueff$. However, for red-giant stars, the theoretical $\taueff$  systematically underestimates  the measurements  by about 40\,\%. This  discrepancy  is of the same order as that obtained  by \citet{Mathur11}. 
Our theoretical $\taueff$ scales as $\numax^{p}$ with $p=- 0.94$, which agrees with \citet{Mathur11}.   Indeed, their average value of the slope is $p= -0.89 $ with a maximum difference of 0.03  between the different methods.   In both cases, however, $\taueff$  significantly departs from a linear scaling  with $\numax$.   






Second, to compare the theoretical scaling relation in $(\numax \, \Ma )^{-1}$ (\eq{scal_taueff2}) with the observations, we used the scaling relation  found for $\Ma$ (\eq{Ma3D}).  This requires knowing $\teff$ and the gravity $g$.    For RG stars, the effective temperatures of the targets were obtained from the {\it Kepler} Input Catalogue \citep{Brown11} and corrected following  \citet{Thygesen12}.
For 136 of 141  sub-giant and MS targets, the effective temperatures  were obtained from the Sloan Digital Sky Survey \citep{Pinsonneault12}. For  the five remaining targets, they were extracted from \cite{Silva-Aguirre12}. 
 The surface gravity $g$ of the targets were determined using  the scaling relation for $\numax$ (\eq{numax}).

We have plotted in Fig.~\ref{fig:taueff_numax_Mascaling} (bottom panel)  measured and theoretical values of $\taueff$ as a function of the quantity $ z_2$ given by \eq{scal_taueff2}. 
The measured values of $\taueff$ are found to scale as $z_2^m$ with $m=1.01$, which  is very close to the expected theoretical scaling in $z_2 \propto (\numax \, \Ma )^{-1}$ and the departure from a linear scaling is lower than that  observed with the classical scaling (i.e. with $c_2 \propto \numax^{-1}$ ). 
However,  the theoretical $\taueff$ systematically underestimates  the observations  by about 40\,\%. 
A large part of this systematic difference is  a consequence of our choice of fitting the granulation  background with a  Lorentzian function (\eq{lorenztian}).



 The new theoretical scaling for $\taueff$ reproduces  the observations on a global scale. However, the question remains  whether or not the observations allow one to quantitatively confirm the  dependence on $\Ma$.  We investigate this question  in Appendix~\ref{diff_taueff} and conclude that the {\it Kepler} observations cannot distinguish the new scaling relation from the classical one.

\subsection{Brightness fluctuation  $\sigma$}
\label{comparision_sigma}



\begin{figure}[ht]
        \centering
         \includegraphics[width=0.98\hsize]  {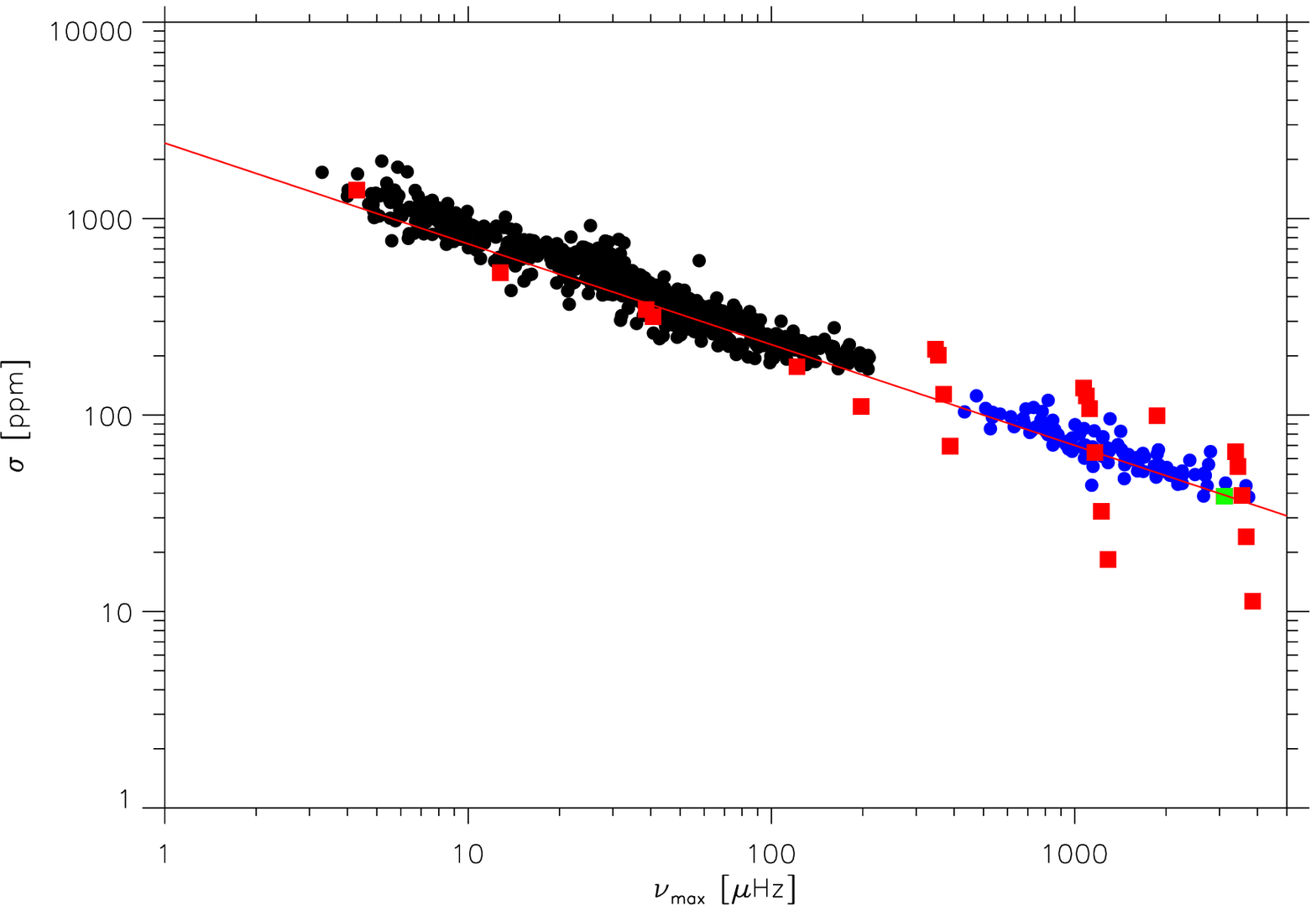}
         \includegraphics[width=0.98\hsize]  {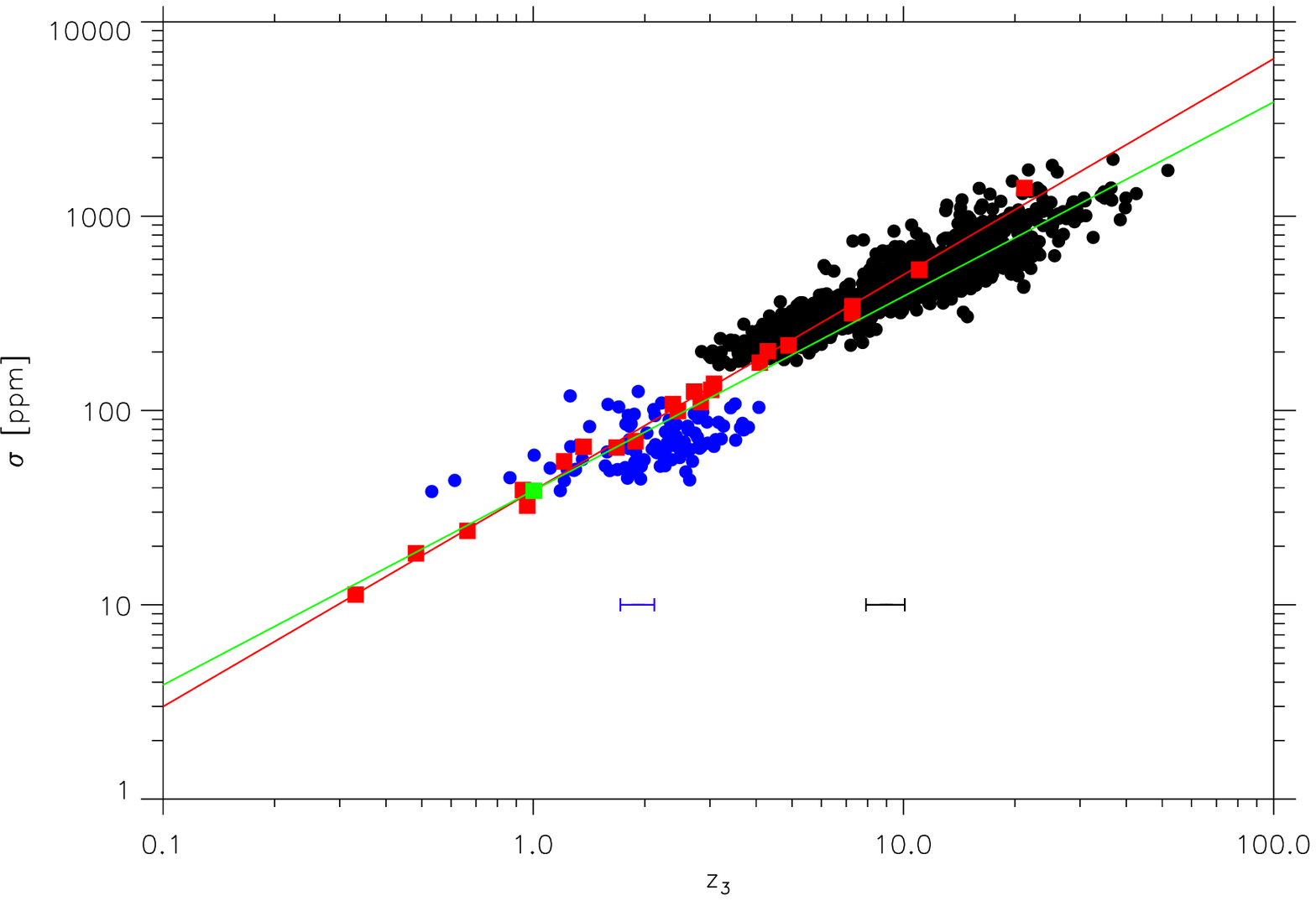}
\caption{{\bf Top:} Root-mean-square brightness fluctuation $\sigma$ as a function of $\numax$. The symbols have the same meaning as in Fig.~\ref{HR}. The red curve is a power law of the form $\numax^{p}$ with the slope $p=-0.51$ obtained by fitting the power law to the theoretical values of $\sigma$ (filled red squares). The filled green square corresponds to the value $\sigma = 39$~ppm found for the solar 3D model. {\bf Bottom:} $\sigma$ as a function of the quantity $z_3$   given by \eq{sigma_scaling_2}  where  $\Ma$ and $\teff^{3/4} \, M^{-1/2}$ are  here supposed to scale according to \eq{Ma3D} and  \eq{scaling_teff_M}, respectively.  The green line corresponds to a linear scaling with $z_3$  and the red one to a power law of the form  $ z_3^{p}$ where $p = 1.11 $.} 
\label{fig:sigma_z_Mascaling}

 \end{figure}

The theoretical values of $\sigma$ are compared in Fig.~\ref{fig:sigma_z_Mascaling} (top panel) with the measured ones as a function of $\numax$. As in \citet{Mathur11}, our theoretical values of $\sigma$  scale approximately as $\numax^{-1/2}$, in agreement with the observations. 
For RG stars,  theoretical $\sigma$  falls within the observational domain. However, 
the  dispersion  in  the theoretical  calculations is much lower than in the measurements. This dispersion in the measurements must be linked to the fact that we observed a sample of stars inhomogeneous in terms of surface metal abundance.



 For sub-giant and MS stars   the observations and the theoretical values overlap. However,  at fixed values of $\numax$, theoretical values of $\sigma$ extend over a wider range than the observations. This is in part because the theoretical calculations include 3D models corresponding to dwarf stars cooler than the observed targets. Indeed, as seen in Fig.~\ref{HR}, sub-giant and MS   stars cooler than about 5\,000~K and with  $\numax \gtrsim ~800~\mu$Hz are lacking in our sample. According to our calculations, these stars are expected to have a value of $\sigma$ about one to two orders of magnitude lower than observed. 



We now compare the theoretical scaling given by \eq{sigma_scaling_2} with the observations.
The Mach number is estimated using \eq{Ma3D},  while the  term $\teff^{3/4} \, M^{-1/2}$ can be derived   using the scaling relations associated with $\numax$ and  the  large separation $\Delta \nu$ \citep[see e.g.][for a recent review see \citet{Kevin13b}]{Stello09,Kallinger10,Mosser10}. 
Indeed, combining the two seismic relations yields
\begin{equation}
{\teff^{3/4} \over M^{1/2}} =  \left ( { \numax \over \numaxref }  \right )^{-3/2} \,  \left ( { \Delta \nu \over \deltanuref }  \right )^{2} \, { \teffsun^{3/4} \over M_\odot ^{1/2} } \; . 
\label{scaling_teff_M}
\end{equation}


In Fig.~\ref{fig:sigma_z_Mascaling} (bottom panel), we have plotted  the measured values $\sigma$ as a function of $z_3$ (\eq{sigma_scaling_2})  where  $\Ma$ and the term $\teff^{3/4} \, M^{-1/2}$ are here evaluated according to the scaling relations given by \eq{Ma3D} and  \eq{scaling_teff_M}, respectively.  The measured values of $\sigma$ are generally aligned with the linear relation in $z_3$. 

 Similarly to the procedure adopted for the scaling relation for $\taueff$, we investigate in Appendix \ref{diff_sigma} the question whether or not the  current observations  can distinguish the new scaling relation  from the classical one.  It is found that compared with the new scaling relation, the classical one results in a smaller difference with the observations.  However,  the deviations of both scaling relations  from the measurements  are found to depend on $\teff$, and the highest deviations are obtained for the F-dwarf stars ($\teff$=6 000~K - 7 500~K). As discussed in Sect.~\ref{discussion}, this is very  likely a consequence of the lack of modelling of the  impact of magnetic activity on the granulation background. Indeed,  a high level of magnetic-activity is expected to inhibit the surface convection and consequently reduce the Mach number. 
 As stressed in \PI, our theoretical calculations must be rigorously valid for  stars with a low level of activity. If we now exclude  the F-dwarf stars from our sample,  we find that the current observations do not allow us to  distinguish the new theoretical scaling relation from the classical one.  
  It is finally established  that only cool  K-dwarf stars will allow us in principle to distinguish  the dependence of $\sigma$ on $\Ma$.

\section{Activity and granulation background}

\label{discussion}

\citet{Chaplin11b} have shown clear evidence that a high level of magnetic activity inhibits the amplitudes of the solar-like oscillations and the strongest effects were --~ on average ~--  observed  for F-dwarfs  \citep{Chaplin11}. The authors concluded that this constitutes   strong evidence for the impact of the magnetic activity on the near-surface convection. In this context, the case of the F-dwarf star HD~49933 is particularly enlightening. Indeed, this star shows clear evidence of a high level of activity \citep[see][]{Mosser05,Mosser09b,Garcia10}. \citet{Ludwig09} have compared the theoretical granulation spectrum computed on the basis of the {\it ab~initio} approach \citep{Ludwig09} with the one measured with CoRoT on HD~49933. 
Their theoretical calculation  results in an  overestimation of the measured $\sigma$ by about 70\,\%, however.   The authors argued that such a discrepancy is  common to the F-dwarfs observed by CoRoT. 

Consistently with the \citet{Ludwig09} results, our theoretical calculations result in a trend towards a high overestimation of the measured $\sigma$ and a moderate underestimation of $\taueff$  for F-dwarf stars. This is illustrated for $\sigma$ in Fig.~\ref{diff:temp}, where we have plotted  the relative differences $D_\sigma$ and $D_\sigma^\prime$ as a function of $\teff$ (see Appendix~\ref{diff_sigma}). 
The strongest relative differences  are obtained around $\teff \approx $ 6 400~K.  On the other hand, the weakest differences are observed on average with  stars cooler than about 6~000~K. 
This observed trend is very likely due to the impact of the magnetic-activity on the granulation background, which is not included in the current modellings. 
Indeed,  a high level of magnetic activity can inhibit the surface convection   to some extent \citep[see e.g.][and references therein]{Nordlund09} and consequently inhibit the granulation background. 

We note that reducting the photospheric Mach number of the F-dwarf stars by about 30\,\%  leads on average to values of $\sigma$ and $\taueff$ close to the  measurements. Reduced thus, the Mach number is at about the same level as that of the solar 3D model ($\Ma  \simeq 0.27$). 
 For instance, the unexpectedly low observed amplitude  measured for   HD~49933  can been explained if we adopt for  this star  $\Ma = 0.30$, which is about 30\,\% lower than predicted for the F-dwarf 3D model representative for HD~49933.  However, whether or not magnetic activity can indeed reduce  $\Ma$ by this amount remains an open question.

\fig{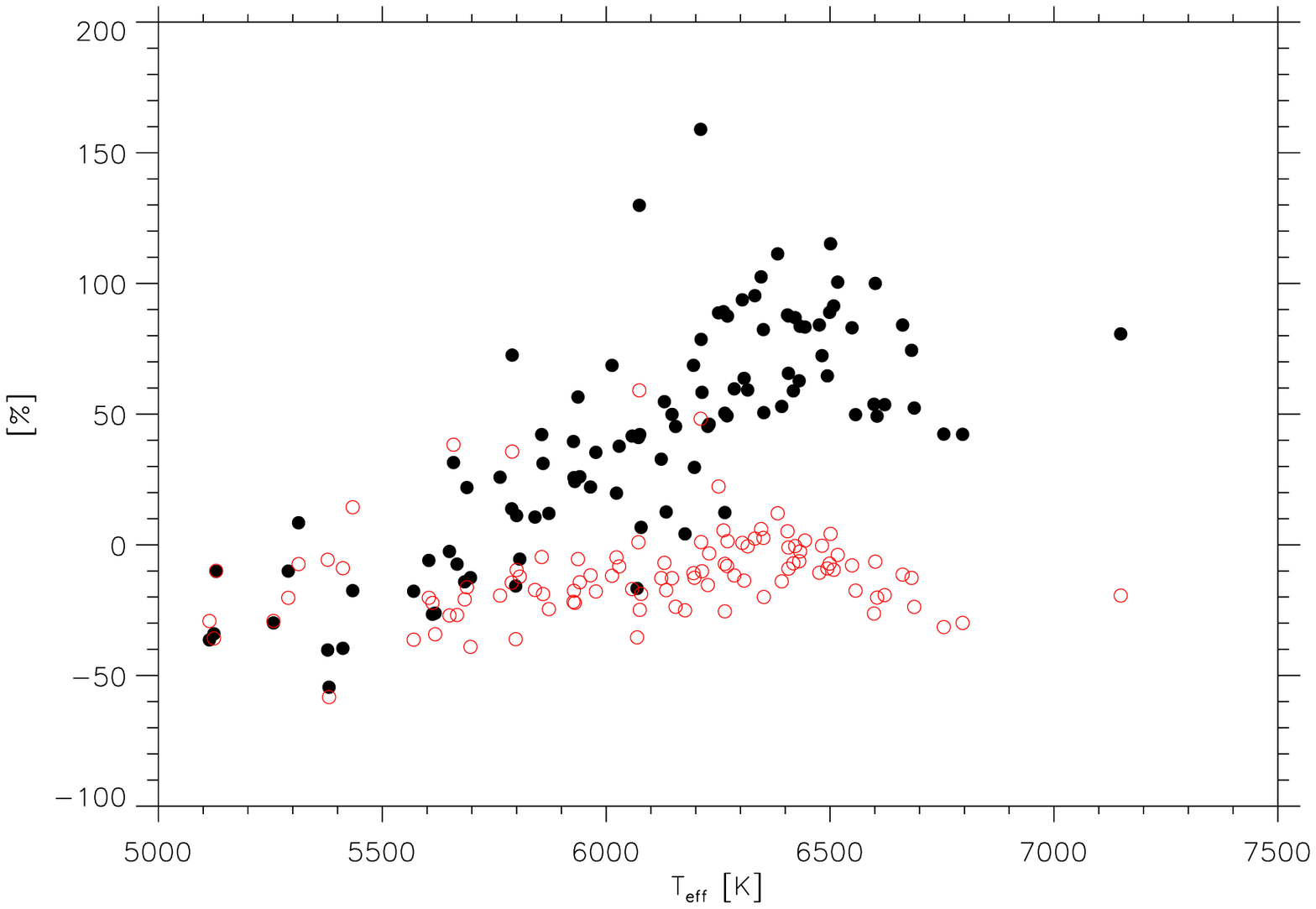}{Relative differences (in \%) between the theoretical $\sigma$ and the measured ones as a function of $\teff$. The black filled circles  correspond to the relative difference $D_{\sigma} = \left (\sigma_\odot /\sigma_m  \right )  \,  z_3 - 1$  and  the red open circles to the relative difference $D_{\sigma}^\prime = \left (\sigma_\odot / \sigma_m \right )  \,   c_3 - 1 $, where $z_3$  (\eq{sigma_scaling_2})  is the new  theoretical scaling relation,  $c_3$ (\eq{c_3}) is the classical one, and $\sigma_m$ the measurements (see details in Appendix~\ref{diff_sigma}).}{diff:temp}

The classical scaling relation also shows an excess around  $\teff \approx $ 6 400~K. 
However, this  is much weaker than for the new  scaling relation. 
Given the fact that the classical scaling relation does obviously not take into account the impact of the magnetic activity, it is  very surprising to observe such a weak excess. However, following our comment above, we believe  that this is a consequence of reduction of $\Ma$ by the  magnetic activity, which led  to values of $\Ma$ for the active F-dwarfs comparable with what is expected for G-dwarfs stars. This would  explain why   $\sigma$ is mainly controlled  by the term $\teff^{3/4}\, M^{-1/2} \,\numax^{-1/2}$, from G-dwarfs to F-dwarfs,  that is, the classical scaling relation.

The discrepancies observed for F-dwarfs stars must  be explained by missing physical processes that involve  the magnetic field at some level.
To test the effect of a local magnetic field on the granulation background, 
\citet{Ludwig09} have computed a set of 2D MHD solar models with different magnetic flux levels. However, a negligible effect on the temporal power spectra of the  emergent intensity  was  obtained, which leads  the authors to conclude that locally generated magnetic fields are unlikely to be responsible for the discrepancy with the observations. We are then left with an enigmatic discrepancy.

\section{Summary and concluding remarks}

\label{summary}

Using a grid of 3D models of the surface layers of various stars, we have computed  the  theoretical power density spectra associated with the granulation background on the basis of the theoretical model presented in \PI. For each  theoretical PDS we  derived a  characteristic time $\taueff$ and amplitude $\sigma$  associated with the granulation background. We  compared these values with the theoretical scaling relations and  observations.

From the current theoretical model we  derived theoretical scaling relations for   $\taueff$ and $\sigma$: 
$\taueff$ was found to scale inversely as $ z_2 \propto (\Ma \,  \numax)$, while  $\sigma$  was found to scale as  $z_3 \propto (\teff^{3/4}/M^{-1/2}) \, \numax^{-1/2} \, f^2(\Ma) $, where $\Ma$ is the turbulent Mach number at the photosphere and $f(\Ma)$ a second-order polynomial function (\eq{theta_scaling}). 
Both scaling relations were found to approximately agree with the individual values derived from the  theoretical PDS obtained with our grid of 3D models.

  The scaling relations predicted by the model depend on $\numax$ in the same way as the classical theoretical  scaling relations. 
 Our  model thus  provides theoretical support for the scaling relation for $\taueff$ that was up to now explained by assuming that the granules move proportionally to the  sound speed $c_s$ \citep{Huber09,Kjeldsen11}. 
Furthermore, consistently with  the prediction by \citet{Ludwig06}, our theoretical scaling relation for $\sigma$ is found to scale as the inverse of the square root of the number of granules over the stellar surface, which is expected to scale as $\numax \, M / \teff^{3/2}$.

The model also predicts that  $\taueff$ and $\sigma$ not only depend on $\numax$, but are also controlled by  the Mach number $\Ma$.   For $\taueff$ this dependence on $\Ma$ is simply explained by  the fact that the granule life-time is ultimately controlled by the ratio between the granule size $\Lambda$ and its  velocity $V$. In turn, this ratio is proportional to the inverse of the product $\numax \, \Ma$, where $\Ma \propto V / c_s $. For red giants, the observed dependence of $\taueff$ with $\numax$ is thus explained by the fact that, during the evolution, variations of $\numax$ dominate the variation of $\Ma$.
The dependence of $\sigma$ with $\Ma$ is  due to the close link between $\sigma$ and the temperature fluctuations $\theta_{\rm rms}$. In turn, there is a balance between $\theta_{\rm rms}$  and the granule kinetic energy (see \eq{w_theta}), and  consequently  between $\theta_{\rm rms}$  and $\Ma$.

To compare these theoretical scaling relations with the observations we have derived  a scaling relation for $\Ma$ using our grid of 3D models. This scaling  has the form $\Ma \propto \teff^{2.4} \, g^{-0.15}$. We  then compared the  theoretical scaling relations for   $\taueff$ and $\sigma$ with the measured values.

 Quantitatively,  the theoretical values $\taueff$  systematically underestimate   the measurements by about 40\,\%.    
This systematic difference  is mainly a consequence of the way $\taueff$ is determined from the granulation spectrum (see Sect.~\ref{comparision_taueff}). 
 Calibrating the theoretical $\taueff$ to the solar reference   results in a difference with the observations of less than 10\,\% on average.
At fixed values of $z_3$, theoretical values of  $\sigma$  systematically  overestimate  the observations made for red giants by about  12\,\%. For a large part, this overestimation was shown to be the consequence of the considerable degeneracy that occurs for red giants  between the mass and radius (see Appendix~\ref{degeneracy}). Comparing instead the quantity $\tilde{\sigma} \equiv (R_s/R_\odot) \, \sigma$ removes the dependence with the masses and radii attributed to the 3D models  and results on average in a difference  with the  measurements of only 5\,\%.
For  MS and sub-giants, the  theoretical $\sigma$  differ on average by about 50\.\% from the measurements. 
This departure is found to depend  on $\teff$. The highest deviations are obtained with the F-dwarf stars and are very likely due to the fact that the impact of the magnetic activity on the granulation background is not modelled (see the discussion in Sect.~\ref{discussion}). If we exclude the F-dwarf stars from our sample,   theoretical $\sigma$ underestimate the observations by only about 2\,\%.

For RG stars, the differences between our predictions for $\sigma$ and $\taueff$ and the measurements  are found to be lower than the differences obtained with the different methods of analysis investigated in  \citet{Mathur11}. For $\sigma$, these differences  remain  much lower than those obtained by \citet{Mathur11}. Indeed,  theoretical  calculations of these authors overestimate the measured $\sigma$ by a factor of about four.  
In view of our results, the \citet{Mathur11} results are surprising since their theoretical calculations are based on the {\it ab~initio}  modelling of   \citet{Ludwig06} and, as shown in \PI\,, our theoretical  1D modelling agrees well with the theoretical PDS computed on the basis of the {\it ab~initio}  modelling for a red giant 3D model \citep{Ludwig11}.


%

The theoretical scaling relations derived from our model  match  the  variations of $\taueff$ and $\sigma$ measured across the HR diagram 
 with {\it Kepler}  data on a global scale. 
Nevertheless, the differences between the new scaling relations and the classical ones are found to be of the same order as the dispersions of the new scaling relations w.r.t. the measurements. It is therefore not possible with our sample of stars to confirm or refute the dependence of $\sigma$ and $\taueff$  with  $\Ma$. 
 The high levels of  the dispersions between the scaling relations and the measurements have two origins. First, $\Ma$ mainly depends on $\teff$, and the precision with which $\teff$ is measured significantly contributes to the dispersion. Second, we observed population of stars inhomogeneous in terms of surface metal abundance. However, the granulation background is expected to depend on the surface metal  abundance  in a manner that remains to be investigated across the HR diagram \citep[for a particular low-metal F-dwarf star see][]{Ludwig09}.

 Detection of solar-like oscillations in a statistically sufficient number of  K-dwarf stars ($\teff= 3\,500 - 5\,000$)  would in principle permit us to test the dependence of $\taueff$ and $\sigma$ on $\Ma$. Indeed, these cool dwarf stars have  $\numax \gtrsim ~800~\mu$Hz ($\log g \gtrsim 4.0$) and   are expected to have simultaneously a significantly lower $\Ma$ and a  $\numax$ comparable to the MS stars for which solar-like oscillations were so far detected with {\it Kepler}. Unfortunately,  such  K-dwarf stars  are lacking in our sample.

In \citet{Chaplin11}, about 760 stars showing solar-like oscillations and observed with a short cadence were flagged by the detection threshold.
Of these targets, those with $\numax \gtrsim$ 0.8~mHz all have  an effective temperature higher than 5~000 K. 
It is therefore very unlikely  to extend the current samples such as to have a statistically sufficient number of K-dwarf stars showing solar-like oscillations.
If we can   verify the  dependence of the scaling relations on  $\Ma$, this will constitute  a confirmation of the theoretical scaling relations derived in the present work. It will also constitute a  way to test for a variety of stellar  standard 1D models of the surface convection.

%

\begin{acknowledgements}
Funding for this Discovery mission is provided by NASA's
Science Mission Directorate. The authors thank the entire {\it Kepler} 
team, without whom these results would not be possible.
 RS, KB, BM and AB acknowledge financial support from the Programme National de Physique Stellaire (PNPS) of CNRS/INSU and from Agence Nationale de la Recherche (ANR, France) program  ``Interaction Des \'Etoiles et des Exoplan\`etes'' (IDEE, ANR-12-BS05-0008). EC and HGL acknowledge financial support by the Sonderforschungsbereich SFB\,881 ``The Milky Way System'' (subproject A4) of the German Research Foundation (DFG). The National Center for Atmospheric Research (NCAR) is partially funded by the National Science Foundation. SM's work was partially supported by NASA grant NNX12AE17G.
\end{acknowledgements}

\bibliographystyle{aa}

\begin{thebibliography}{59}
\expandafter\ifx\csname natexlab\endcsname\relax\def\natexlab#1{#1}\fi

\bibitem[{{Aigrain} {et~al.}(2004){Aigrain}, {Favata}, \&
  {Gilmore}}]{Aigrain04}
{Aigrain}, S., {Favata}, F., \& {Gilmore}, G. 2004, in ESA Special Publication,
  Vol. 538, Stellar Structure and Habitable Planet Finding, ed. F.~{Favata},
  S.~{Aigrain}, \& A.~{Wilson}, 215--224

\bibitem[{{Andersen} {et~al.}(1998){Andersen}, {Leifsen}, {Appourchaux},
  {Frohlich}, {Jim{\'e}nez}, \& {Wehrli}}]{Andersen98}
{Andersen}, B., {Leifsen}, T., {Appourchaux}, T., {et~al.} 1998, in ESA Special
  Publication, Vol. 418, Structure and Dynamics of the Interior of the Sun and
  Sun-like Stars, ed. {S.~Korzennik}, 83

\bibitem[{{Asplund} {et~al.}(2005){Asplund}, {Grevesse}, \&
  {Sauval}}]{Asplund05}
{Asplund}, M., {Grevesse}, N., \& {Sauval}, A.~J. 2005, in Astronomical Society
  of the Pacific Conference Series, Vol. 336, Cosmic Abundances as Records of
  Stellar Evolution and Nucleosynthesis, ed. T.~G. {Barnes}, III \& F.~N.
  {Bash}, 25

\bibitem[{{Ballot} {et~al.}(2011){Ballot}, {Barban}, \& {van't
  Veer-Menneret}}]{Ballot11b}
{Ballot}, J., {Barban}, C., \& {van't Veer-Menneret}, C. 2011, \aap, 531, A124

\bibitem[{{Bedding} \& {Kjeldsen}(2003)}]{Bedding03}
{Bedding}, T.~R. \& {Kjeldsen}, H. 2003, Publications of the Astronomical
  Society of Australia, 20, 203

\bibitem[{{Belkacem} {et~al.}(2011){Belkacem}, {Goupil}, {Dupret}, {Samadi},
  {Baudin}, {Noels}, \& {Mosser}}]{Kevin11}
{Belkacem}, K., {Goupil}, M.~J., {Dupret}, M.~A., {et~al.} 2011, \aap, 530,
  A142

\bibitem[{{Belkacem} {et~al.}(2013){Belkacem}, {Samadi}, {Mosser}, {Goupil}, \&
  {Ludwig}}]{Kevin13b}
{Belkacem}, K., {Samadi}, R., {Mosser}, B., {Goupil}, M.-J., \& {Ludwig}, H.-G.
  2013, in Progress in physics of the sun and stars: a new era in helio- and
  asteroseismology, Vol. to appera in ASP Conference Series [arXiv:1307.3132]

\bibitem[{{Brown} {et~al.}(1991){Brown}, {Gilliland}, {Noyes}, \&
  {Ramsey}}]{Brown91}
{Brown}, T.~M., {Gilliland}, R.~L., {Noyes}, R.~W., \& {Ramsey}, L.~W. 1991,
  \apj, 368, 599

\bibitem[{{Brown} {et~al.}(2011){Brown}, {Latham}, {Everett}, \&
  {Esquerdo}}]{Brown11}
{Brown}, T.~M., {Latham}, D.~W., {Everett}, M.~E., \& {Esquerdo}, G.~A. 2011,
  \aj, 142, 112

\bibitem[{{Bruntt} {et~al.}(2012){Bruntt}, {Basu}, {Smalley}, {Chaplin},
  {Verner}, {Bedding}, {Catala}, {Gazzano}, {Molenda-{\.Z}akowicz}, {Thygesen},
  {Uytterhoeven}, {Hekker}, {Huber}, {Karoff}, {Mathur}, {Mosser},
  {Appourchaux}, {Campante}, {Elsworth}, {Garc{\'{\i}}a}, {Handberg},
  {Metcalfe}, {Quirion}, {R{\'e}gulo}, {Roxburgh}, {Stello},
  {Christensen-Dalsgaard}, {Kawaler}, {Kjeldsen}, {Morris}, {Quintana}, \&
  {Sanderfer}}]{Bruntt12}
{Bruntt}, H., {Basu}, S., {Smalley}, B., {et~al.} 2012, \mnras, 423, 122

\bibitem[{{Bruntt} {et~al.}(2011){Bruntt}, {Frandsen}, \&
  {Thygesen}}]{Bruntt11}
{Bruntt}, H., {Frandsen}, S., \& {Thygesen}, A.~O. 2011, \aap, 528, A121

\bibitem[{{Chaplin} {et~al.}(2011{\natexlab{a}}){Chaplin}, {Bedding},
  {Bonanno}, {Broomhall}, {Garc{\'{\i}}a}, {Hekker}, {Huber}, {Verner}, {Basu},
  {Elsworth}, {Houdek}, {Mathur}, {Mosser}, {New}, {Stevens}, {Appourchaux},
  {Karoff}, {Metcalfe}, {Molenda-{\.Z}akowicz}, {Monteiro}, {Thompson},
  {Christensen-Dalsgaard}, {Gilliland}, {Kawaler}, {Kjeldsen}, {Ballot},
  {Benomar}, {Corsaro}, {Campante}, {Gaulme}, {Hale}, {Handberg}, {Jarvis},
  {R{\'e}gulo}, {Roxburgh}, {Salabert}, {Stello}, {Mullally}, {Li}, \&
  {Wohler}}]{Chaplin11b}
{Chaplin}, W.~J., {Bedding}, T.~R., {Bonanno}, A., {et~al.} 2011{\natexlab{a}},
  \apjl, 732, L5

\bibitem[{{Chaplin} {et~al.}(2011{\natexlab{b}}){Chaplin}, {Kjeldsen},
  {Bedding}, {Christensen-Dalsgaard}, {Gilliland}, {Kawaler}, {Appourchaux},
  {Elsworth}, {Garc{\'{\i}}a}, {Houdek}, {Karoff}, {Metcalfe},
  {Molenda-{\.Z}akowicz}, {Monteiro}, {Thompson}, {Verner}, {Batalha},
  {Borucki}, {Brown}, {Bryson}, {Christiansen}, {Clarke}, {Jenkins}, {Klaus},
  {Koch}, {An}, {Ballot}, {Basu}, {Benomar}, {Bonanno}, {Broomhall},
  {Campante}, {Corsaro}, {Creevey}, {Esch}, {Gai}, {Gaulme}, {Hale},
  {Handberg}, {Hekker}, {Huber}, {Mathur}, {Mosser}, {New}, {Pinsonneault},
  {Pricopi}, {Quirion}, {R{\'e}gulo}, {Roxburgh}, {Salabert}, {Stello}, \&
  {Suran}}]{Chaplin11}
{Chaplin}, W.~J., {Kjeldsen}, H., {Bedding}, T.~R., {et~al.}
  2011{\natexlab{b}}, \apj, 732, 54

\bibitem[{Cox(1968)}]{Cox68}
Cox, J. 1968, Principles of stellar structure (Gordon and Breach)

\bibitem[{{Freytag} {et~al.}(2012){Freytag}, {Steffen}, {Ludwig},
  {Wedemeyer-B{\"o}hm}, {Schaffenberger}, \& {Steiner}}]{Freytag12}
{Freytag}, B., {Steffen}, M., {Ludwig}, H.-G., {et~al.} 2012, Journal of
  Computational Physics, 231, 919

\bibitem[{{Garc{\'{\i}}a} {et~al.}(2011){Garc{\'{\i}}a}, {Hekker}, {Stello},
  {Guti{\'e}rrez-Soto}, {Handberg}, {Huber}, {Karoff}, {Uytterhoeven},
  {Appourchaux}, {Chaplin}, {Elsworth}, {Mathur}, {Ballot},
  {Christensen-Dalsgaard}, {Gilliland}, {Houdek}, {Jenkins}, {Kjeldsen},
  {McCauliff}, {Metcalfe}, {Middour}, {Molenda-Zakowicz}, {Monteiro}, {Smith},
  \& {Thompson}}]{Garcia11}
{Garc{\'{\i}}a}, R.~A., {Hekker}, S., {Stello}, D., {et~al.} 2011, \mnras, 414,
  L6

\bibitem[{{Garc{\'{\i}}a} {et~al.}(2010){Garc{\'{\i}}a}, {Mathur}, {Salabert},
  {Ballot}, {R{\'e}gulo}, {Metcalfe}, \& {Baglin}}]{Garcia10}
{Garc{\'{\i}}a}, R.~A., {Mathur}, S., {Salabert}, D., {et~al.} 2010, Science,
  329, 1032

\bibitem[{{Guenther} {et~al.}(2008){Guenther}, {Kallinger}, {Gruberbauer},
  {Huber}, {Weiss}, {Kuschnig}, {Demarque}, {Robinson}, {Matthews}, {Moffat},
  {Rucinski}, {Sasselov}, \& {Walker}}]{Guenther08}
{Guenther}, D.~B., {Kallinger}, T., {Gruberbauer}, M., {et~al.} 2008, \apj,
  687, 1448

\bibitem[{{Hansen} \& {Kawaler}(1994)}]{Hansen94}
{Hansen}, C.~J. \& {Kawaler}, S.~D. 1994, {Stellar Interiors. Physical
  Principles, Structure, and Evolution.}

\bibitem[{{Harvey}(1985)}]{Harvey85}
{Harvey}, J. 1985, in ESA Special Publication, Vol. 235, Future Missions in
  Solar, Heliospheric \& Space Plasma Physics, ed. {E.~Rolfe \& B.~Battrick},
  199

\bibitem[{{Harvey} {et~al.}(1993){Harvey}, {Duvall}, {Jefferies}, \&
  {Pomerantz}}]{Harvey93}
{Harvey}, J.~W., {Duvall}, Jr., T.~L., {Jefferies}, S.~M., \& {Pomerantz},
  M.~A. 1993, in Astronomical Society of the Pacific Conference Series,
  Vol.~42, GONG 1992. Seismic Investigation of the Sun and Stars, ed. T.~M.
  {Brown}, 111

\bibitem[{{Hekker} {et~al.}(2011){Hekker}, {Elsworth}, {De Ridder}, {Mosser},
  {Garc{\'{\i}}a}, {Kallinger}, {Mathur}, {Huber}, {Buzasi}, {Preston}, {Hale},
  {Ballot}, {Chaplin}, {R{\'e}gulo}, {Bedding}, {Stello}, {Borucki}, {Koch},
  {Jenkins}, {Allen}, {Gilliland}, {Kjeldsen}, \&
  {Christensen-Dalsgaard}}]{Hekker11b}
{Hekker}, S., {Elsworth}, Y., {De Ridder}, J., {et~al.} 2011, \aap, 525, A131

\bibitem[{{Houdek} {et~al.}(1999){Houdek}, {Balmforth},
  {Christensen-Dalsgaard}, \& {Gough}}]{Houdek99}
{Houdek}, G., {Balmforth}, N.~J., {Christensen-Dalsgaard}, J., \& {Gough},
  D.~O. 1999, \aap, 351, 582

\bibitem[{{Huber} {et~al.}(2009){Huber}, {Stello}, {Bedding}, {Chaplin},
  {Arentoft}, {Quirion}, \& {Kjeldsen}}]{Huber09}
{Huber}, D., {Stello}, D., {Bedding}, T.~R., {et~al.} 2009, Communications in
  Asteroseismology, 160, 74

\bibitem[{{Kallinger} \& {Matthews}(2010)}]{Kallinger10b}
{Kallinger}, T. \& {Matthews}, J.~M. 2010, \apjl, 711, L35

\bibitem[{{Kallinger} {et~al.}(2010){Kallinger}, {Weiss}, {Barban}, {Baudin},
  {Cameron}, {Carrier}, {De Ridder}, {Goupil}, {Gruberbauer}, {Hatzes},
  {Hekker}, {Samadi}, \& {Deleuil}}]{Kallinger10}
{Kallinger}, T., {Weiss}, W.~W., {Barban}, C., {et~al.} 2010, \aap, 509, A77

\bibitem[{{Karoff}(2012)}]{Karoff11}
{Karoff}, C. 2012, \mnras, 421, 3170

\bibitem[{{Karoff} {et~al.}(2013){Karoff}, {Campante}, {Ballot}, {Kallinger},
  {Gruberbauer}, {Garcia}, {Caldwell}, {Christiansen}, \&
  {Kinemuchi}}]{Karoff13}
{Karoff}, C., {Campante}, T.~L., {Ballot}, J., {et~al.} 2013, ArXiv e-prints

\bibitem[{{Kjeldsen} \& {Bedding}(1995)}]{Kjeldsen95}
{Kjeldsen}, H. \& {Bedding}, T.~R. 1995, \aap, 293, 87

\bibitem[{{Kjeldsen} \& {Bedding}(2011)}]{Kjeldsen11}
{Kjeldsen}, H. \& {Bedding}, T.~R. 2011, \aap, 529, L8

\bibitem[{{Ludwig}(2006)}]{Ludwig06}
{Ludwig}, H. 2006, \aap, 445, 661

\bibitem[{{Ludwig} {et~al.}(2009{\natexlab{a}}){Ludwig}, {Samadi}, {Steffen},
  {Appourchaux}, {Baudin}, {Belkacem}, {Boumier}, {Goupil}, \&
  {Michel}}]{Ludwig09}
{Ludwig}, H., {Samadi}, R., {Steffen}, M., {et~al.} 2009{\natexlab{a}}, \aap,
  506, 167

\bibitem[{{Ludwig} {et~al.}(2009{\natexlab{b}}){Ludwig}, {Caffau}, {Steffen},
  {Freytag}, {Bonifacio}, \& {Ku{\v c}inskas}}]{Ludwig09b}
{Ludwig}, H.-G., {Caffau}, E., {Steffen}, M., {et~al.} 2009{\natexlab{b}},
  \memsai, 80, 711

\bibitem[{{Ludwig} \& {Steffen}(2011)}]{Ludwig11}
{Ludwig}, H.-G. \& {Steffen}, M. 2011, ArXiv e-prints

\bibitem[{{Mathur} {et~al.}(2011){Mathur}, {Hekker}, {Trampedach}, {Ballot},
  {Kallinger}, {Buzasi}, {Garc{\'{\i}}a}, {Huber}, {Jim{\'e}nez}, {Mosser},
  {Bedding}, {Elsworth}, {R{\'e}gulo}, {Stello}, {Chaplin}, {De Ridder},
  {Hale}, {Kinemuchi}, {Kjeldsen}, {Mullally}, \& {Thompson}}]{Mathur11}
{Mathur}, S., {Hekker}, S., {Trampedach}, R., {et~al.} 2011, \apj, 741, 119

\bibitem[{{Michel} {et~al.}(2008){Michel}, {Baglin}, {Auvergne}, {Catala},
  {Samadi}, {Baudin}, {Appourchaux}, {Barban}, {Weiss}, {Berthomieu},
  {Boumier}, {Dupret}, {Garcia}, {Fridlund}, {Garrido}, {Goupil}, {Kjeldsen},
  {Lebreton}, {Mosser}, {Grotsch-Noels}, {Janot-Pacheco}, {Provost},
  {Roxburgh}, {Thoul}, {Toutain}, {Tiphene}, {Turck-Chieze}, {Vauclair},
  {Aerts}, {Alecian}, {Ballot}, {Charpinet}, {Hubert}, {Lignieres}, {Mathias},
  {Monteiro}, {Neiner}, \& {Poretti}}]{Michel08}
{Michel}, E., {Baglin}, A., {Auvergne}, M., {et~al.} 2008, Science, 322, 558

\bibitem[{{Michel} {et~al.}(2009){Michel}, {Samadi}, {Baudin}, {Barban},
  {Appourchaux}, \& {Auvergne}}]{Michel09}
{Michel}, E., {Samadi}, R., {Baudin}, F., {et~al.} 2009, \aap, 495, 979

\bibitem[{{Molenda-{\.Z}akowicz} {et~al.}(2010){Molenda-{\.Z}akowicz},
  {Bruntt}, {Sousa}, {Frasca}, {Biazzo}, {Huber}, {Ireland}, {Bedding},
  {Stello}, {Uytterhoeven}, {Dreizler}, {De Cat}, {Briquet}, {Catanzaro},
  {Karoff}, {Frandsen}, \& {Spezzi}}]{Molenda-Zakowicz10}
{Molenda-{\.Z}akowicz}, J., {Bruntt}, H., {Sousa}, S., {et~al.} 2010,
  Astronomische Nachrichten, 331, 981

\bibitem[{{Morel} \& {Lebreton}(2008)}]{Morel08}
{Morel}, P. \& {Lebreton}, Y. 2008, \apss, 316, 61

\bibitem[{{Morel} \& {Miglio}(2012)}]{Morel12}
{Morel}, T. \& {Miglio}, A. 2012, \mnras, 419, L34

\bibitem[{{Mosser} {et~al.}(2009){Mosser}, {Baudin}, {Lanza}, {Hulot},
  {Catala}, {Baglin}, \& {Auvergne}}]{Mosser09b}
{Mosser}, B., {Baudin}, F., {Lanza}, A.~F., {et~al.} 2009, \aap, 506, 245

\bibitem[{{Mosser} {et~al.}(2010){Mosser}, {Belkacem}, {Goupil}, {Miglio},
  {Morel}, {Barban}, {Baudin}, {Hekker}, {Samadi}, {De Ridder}, {Weiss},
  {Auvergne}, \& {Baglin}}]{Mosser10}
{Mosser}, B., {Belkacem}, K., {Goupil}, M.-J., {et~al.} 2010, \aap, 517, A22

\bibitem[{{Mosser} {et~al.}(2005){Mosser}, {Bouchy}, {Catala}, {Michel},
  {Samadi}, {Th{\' e}venin}, {Eggenberger}, {Sosnowska}, {Moutou}, \&
  {Baglin}}]{Mosser05}
{Mosser}, B., {Bouchy}, F., {Catala}, C., {et~al.} 2005, \aap, 431, L13

\bibitem[{{Mosser} {et~al.}(2012){Mosser}, {Elsworth}, {Hekker}, {Huber},
  {Kallinger}, {Mathur}, {Belkacem}, {Goupil}, {Samadi}, {Barban}, {Bedding},
  {Chaplin}, {Garc{\'{\i}}a}, {Stello}, {De Ridder}, {Middour}, {Morris}, \&
  {Quintana}}]{Mosser12}
{Mosser}, B., {Elsworth}, Y., {Hekker}, S., {et~al.} 2012, \aap, 537, A30

\bibitem[{{Mosser} {et~al.}(2013){Mosser}, {Michel}, {Belkacem}, {Goupil},
  {Baglin}, {Barban}, {Provost}, {Samadi}, {Auvergne}, \& {Catala}}]{Mosser13}
{Mosser}, B., {Michel}, E., {Belkacem}, K., {et~al.} 2013, \aap, 550, A126

\bibitem[{{Nordlund} {et~al.}(2009){Nordlund}, {Stein}, \&
  {Asplund}}]{Nordlund09}
{Nordlund}, {\AA}., {Stein}, R.~F., \& {Asplund}, M. 2009, Living Reviews in
  Solar Physics, 6, 2

\bibitem[{{Noyes} {et~al.}(1984){Noyes}, {Hartmann}, {Baliunas}, {Duncan}, \&
  {Vaughan}}]{Noyes84}
{Noyes}, R.~W., {Hartmann}, L.~W., {Baliunas}, S.~L., {Duncan}, D.~K., \&
  {Vaughan}, A.~H. 1984, \apj, 279, 763

\bibitem[{{Pinsonneault} {et~al.}(2012){Pinsonneault}, {An},
  {Molenda-{\.Z}akowicz}, {Chaplin}, {Metcalfe}, \& {Bruntt}}]{Pinsonneault12}
{Pinsonneault}, M.~H., {An}, D., {Molenda-{\.Z}akowicz}, J., {et~al.} 2012,
  \apjs, 199, 30

\bibitem[{{Samadi} {et~al.}(2013){Samadi}, {Belkacem}, \& {Ludwig}}]{Samadi13a}
{Samadi}, R., {Belkacem}, K., \& {Ludwig}, H.-G. 2013, submited to \aap

\bibitem[{{Samadi} {et~al.}(2010{\natexlab{a}}){Samadi}, {Ludwig}, {Belkacem},
  {Goupil}, {Benomar}, {Mosser}, {Dupret}, {Baudin}, {Appourchaux}, \&
  {Michel}}]{Samadi09b}
{Samadi}, R., {Ludwig}, H.-G., {Belkacem}, K., {et~al.} 2010{\natexlab{a}},
  \aap, 509, A16

\bibitem[{{Samadi} {et~al.}(2010{\natexlab{b}}){Samadi}, {Ludwig}, {Belkacem},
  {Goupil}, \& {Dupret}}]{Samadi09a}
{Samadi}, R., {Ludwig}, H.-G., {Belkacem}, K., {Goupil}, M.~J., \& {Dupret},
  M.-A. 2010{\natexlab{b}}, \aap, 509, A15

\bibitem[{{Silva Aguirre} {et~al.}(2012){Silva Aguirre}, {Casagrande}, {Basu},
  {Campante}, {Chaplin}, {Huber}, {Miglio}, {Serenelli}, {Ballot}, {Bedding},
  {Christensen-Dalsgaard}, {Creevey}, {Elsworth}, {Garc{\'{\i}}a}, {Gilliland},
  {Hekker}, {Kjeldsen}, {Mathur}, {Metcalfe}, {Monteiro}, {Mosser},
  {Pinsonneault}, {Stello}, {Weiss}, {Tenenbaum}, {Twicken}, \&
  {Uddin}}]{Silva-Aguirre12}
{Silva Aguirre}, V., {Casagrande}, L., {Basu}, S., {et~al.} 2012, \apj, 757, 99

\bibitem[{{Stello} {et~al.}(2009){Stello}, {Chaplin}, {Basu}, {Elsworth}, \&
  {Bedding}}]{Stello09}
{Stello}, D., {Chaplin}, W.~J., {Basu}, S., {Elsworth}, Y., \& {Bedding}, T.~R.
  2009, \mnras, 400, L80

\bibitem[{{Svensson} \& {Ludwig}(2005)}]{Svensson05}
{Svensson}, F. \& {Ludwig}, H.-G. 2005, in ESA Special Publication, Vol. 560,
  13th Cambridge Workshop on Cool Stars, Stellar Systems and the Sun, ed.
  F.~{Favata}, G.~A.~J. {Hussain}, \& B.~{Battrick}, 979

\bibitem[{{Thygesen} {et~al.}(2012){Thygesen}, {Frandsen}, {Bruntt},
  {Kallinger}, {Andersen}, {Elsworth}, {Hekker}, {Karoff}, {Stello},
  {Brogaard}, {Burke}, {Caldwell}, \& {Christiansen}}]{Thygesen12}
{Thygesen}, A.~O., {Frandsen}, S., {Bruntt}, H., {et~al.} 2012, \aap, 543, A160

\bibitem[{{Trampedach}(2004)}]{Trampedach04}
{Trampedach}, R. 2004, PhD thesis, MICHIGAN STATE UNIVERSITY

\bibitem[{{Trampedach} {et~al.}(1998){Trampedach}, {Christensen-Dalsgaard},
  {Nordlund}, \& {Stein}}]{Trampedach98}
{Trampedach}, R., {Christensen-Dalsgaard}, J., {Nordlund}, A., \& {Stein},
  R.~F. 1998, in The First MONS Workshop: Science with a Small Space Telescope,
  ed. H.~{Kjeldsen} \& T.~R. {Bedding}, 59

\bibitem[{{Tremblay} {et~al.}(2013){Tremblay}, {Ludwig}, {Freytag}, {Steffen},
  \& {Caffau}}]{Tremblay13}
{Tremblay}, P.-E., {Ludwig}, H.-G., {Freytag}, B., {Steffen}, M., \& {Caffau},
  E. 2013, \aap, 557, A7

\bibitem[{{V{\'a}zquez Rami{\'o}} {et~al.}(2005){V{\'a}zquez Rami{\'o}},
  {R{\'e}gulo}, \& {Roca Cort{\'e}s}}]{Vazquez-Ramio05}
{V{\'a}zquez Rami{\'o}}, H., {R{\'e}gulo}, C., \& {Roca Cort{\'e}s}, T. 2005,
  \aap, 443, L11

\end{thebibliography}

%
%

\Online

\appendix

\section{Can we distinguish the dependence on  $\Ma$?}

\label{diff}

 We have seen  that the new theoretical scaling relations for $\taueff$ and $\sigma$ are --~ on a  global scale ~-- aligned  with the {\it Kepler} measurements. 
However, the question we address here is whether or not the observations allow one to quantitatively confirm the dependence of the new scaling relations on $\Ma$. To this end we compare the new scaling relations with the classical ones.

 \subsection{Characteristic time-scale, $\taueff$}
\label{diff_taueff}

To check the dependence of the new theoretical scaling relation for $\taueff$ on  $\Ma$, we computed the relative differences between the  new  scaling relation  and the measurements as well as the relative differences  between the classical theoretical scaling relation $\taueff \propto \numax^{-1}$  and the measurements. 
In practice, we computed the quantities $D_{\tau} = \left (\tau_{{\rm eff},\odot} / \tau_{{\rm eff},m}  \right ) \,  z_2   - 1 $ and $D_{\tau}^\prime = \left (\tau_{{\rm eff},\odot} / \tau_{{\rm eff},m}  \right )  \, c_2     - 1 $, where $\tau_{{\rm eff},m}$ is the measured value of $\taueff$, $\tau_{{\rm eff},\odot} = 230~$s is the adopted solar reference \citep{Michel08},  $c_2 = \left ( \numaxref / \numax \right ) $, and $z_2$   is the new scaling relation (\eq{scal_taueff2}). We considered in our comparison only the sample of  MS and sub-giant stars because they are better indicator for the dependence on $\Ma$. 

\begin{figure}
        \centering
 \includegraphics[width=0.98\hsize]  {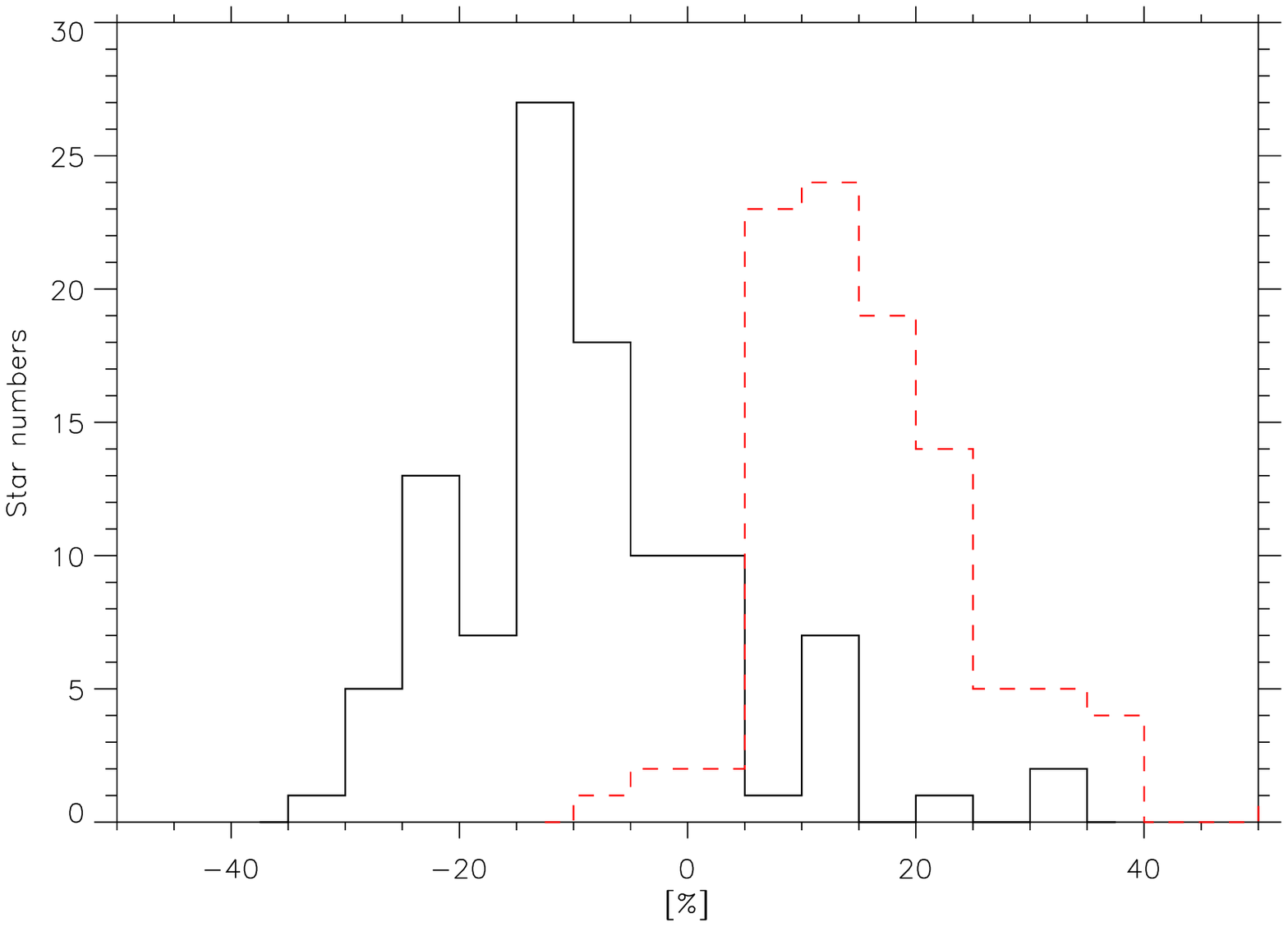}
 \includegraphics[width=0.98\hsize]  {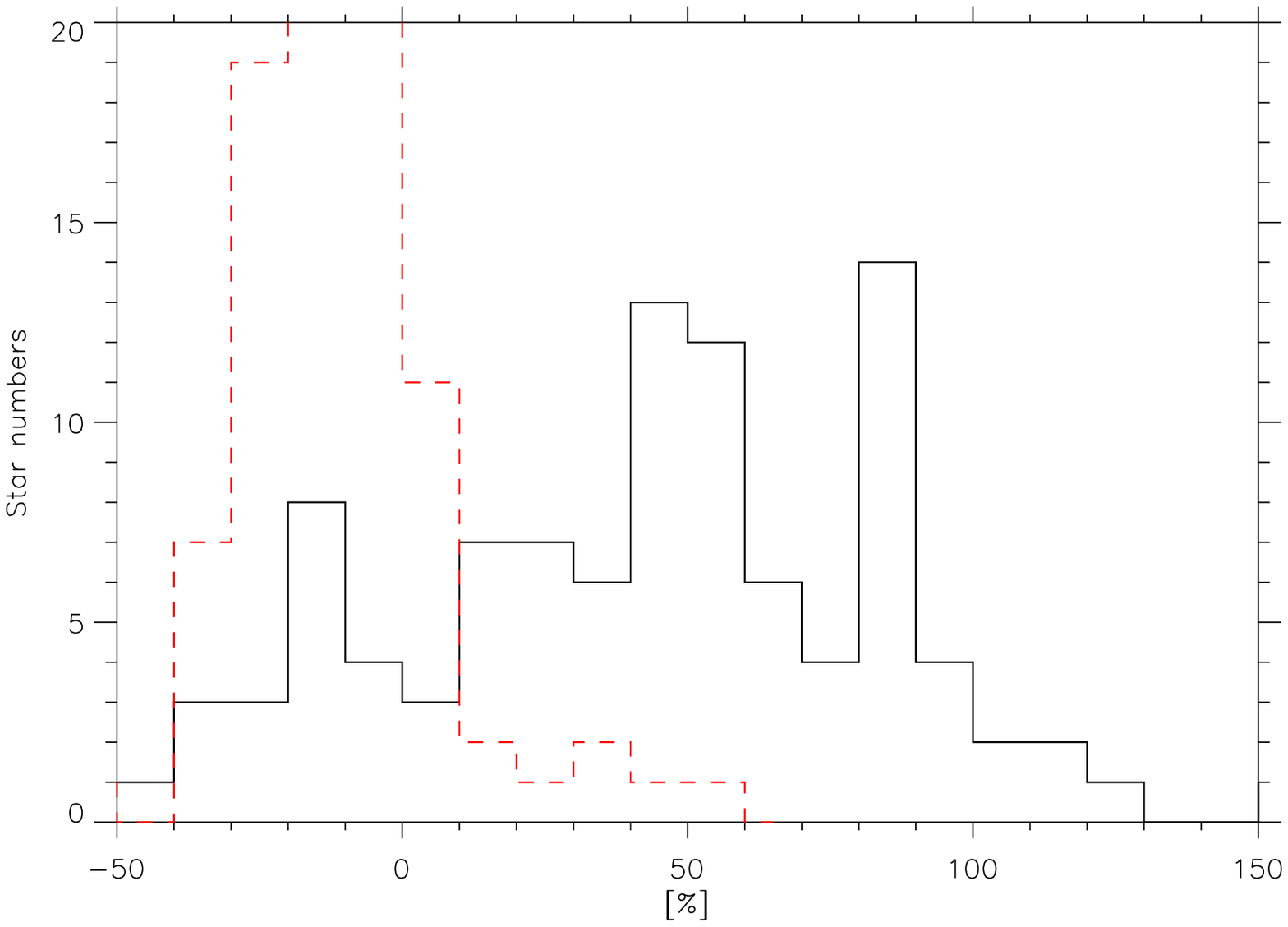}
\caption{ {\bf Top:} Histogram of the relative differences (in \%) between the theoretical $\taueff$ and the measured ones. The solid black line corresponds to the histogram of the relative difference  $D_{\tau} = \left (  \tau_{{\rm eff},\odot} /  \tau_{{\rm eff},m}  \right ) \, z_2  - 1 $  and the dashed red line to  residuals  $D_{\tau}^\prime =  \left (\tau_{{\rm eff},\odot}/ \tau_{{\rm eff},m} \right )  \, c_2    -1 $, where $z_2 =  ( \numaxref/\numax)   \, (\MaO / \Ma )$  is our  theoretical scaling relation and  $ c_2 = \left ( \numaxref / \numax \right )$  the classical one. {\bf Bottom:} Histogram of the relative differences (in \%) between the theoretical $\sigma$ and the measured ones. The solid black line correspond to the histogram of the relative difference $D_{\sigma} = \left (\sigma_\odot /\sigma_m  \right )  \,  z_3 - 1$  and  the dashed red line to the relative difference $D_{\sigma}^\prime = \left (\sigma_\odot / \sigma_m \right )  \,   c_3 - 1 $, where $z_3$  (\eq{sigma_scaling_2})  is the new  theoretical scaling relation and  $c_3$ (\eq{c_3}) is the classical one.
}
\label{fig:residue2}
\end{figure}

The histograms of  $D_{\tau}$ and  $D_{\tau}^\prime$ are shown in Fig.~\ref{fig:residue2} (top panel).
The median value and the standard deviation of $D_{\tau}$ are  $-10$\,\%  and 12\,\%, respectively, while for $D_{\tau}^\prime$  they are equal to  15\,\% and 14\,\%, respectively.

For both scaling relations, the dispersion and deviation from the measurements can in part arise because  we observed an heterogeneous population of stars, in particular stars with different metal abundance. Indeed, $\Ma$ is expected to depend  on the surface metal abundance \citep[see e.g.][]{Houdek99,Samadi09a,Samadi09b}. However, we would have expected a  higher dispersion for  $D_{\tau}$ than for  $D_{\tau}^\prime$. Indeed, the new scaling relation depends on $\Ma$ and, according to \eq{Ma3D}, the Mach number $\Ma$ strongly depends on $\teff$ and more weakly on $g$. Therefore, the uncertainties associated with $\teff$ and $\log g$ introduce  a spread in the determination of $z_2$, and subsequently on $D_{\tau}$.

$\teff$ is  based on  photometric indices and is measured with an rms precision of about 100~K \citep[see][]{Molenda-Zakowicz10,Bruntt11,Thygesen12,Bruntt12},  while $\log g$ is obtained from seismology  with a typical rms precision of 0.1~dex \citep{Bruntt12,Morel12}.  The rms errors   in $\teff$ and $\log g$  introduce a relative dispersion in $z_2$ of the order of 6~\% for a typical RG star with $\teff = 4\,500$~K and $\log g = 2.3$, and about $5$~\% for a typical MS  with $\teff = 6\,000$~K  and $\log g=4$ (these typical relative dispersions are shown in Fig.~\ref{fig:taueff_numax_Mascaling}). 


 The median deviation of the  new theoretical scaling relation from the measurements is  found to be of the same order as that of the classical relation.   However, the difference between the median of $D_{\tau}$ and  of $D_{\tau}^\prime$  remains  within the dispersion of $D_{\tau}$. Therefore, we cannot distinguish the new scaling relation from the classical one. 
Finally, the mean deviation of the new scaling relation from the measurements is about two times lower than its associated dispersion.    We therefore conclude that, as the classical scaling relation,  the new  one is compatible with the observations, but we cannot firmly confirm its dependence with $\Ma$. 

\subsection{Brightness fluctuations,  $\sigma$}
\label{diff_sigma}

In the same way as for  $\taueff$, we checked the dependence with $\Ma$ of the new scaling relation for $\sigma$ by computing the relative difference between the new theoretical scaling relation (\eq{sigma_scaling_2}) and the measurements as well as the  relative difference between the classical theoretical scaling relation $\teff^{3/4} \, M^{-1/2} \, \numax^{-1/2}$  \citep{Kjeldsen11,Mathur11} and the measurements.
In practice, we computed the quantities $D_{\sigma} = \left (\sigma_\odot /\sigma_m  \right )  \,  z_3 - 1$  and $D_{\sigma}^\prime = \left (\sigma_\odot / \sigma_m \right )  \,   c_3 - 1 $, where $\sigma_m$ is the measured value of $c_2 \propto \sigma$, $\sigma_\odot = 43~$ppm is the adopted bolometric amplitude measured for the Sun \citep{Michel08}, and $c_3$ is given by \eq{c_3}, where the term  $\left ( {\teff / \teffsun} \right ) ^{3/4} \, \left ( { M_\odot /  M} \right )^{1/2} $ is evaluated according to \eq{scaling_teff_M}.
Like for  $\taueff$, we considered only  MS and sub-giant stars because they are the better indicators.

We have plotted in Fig.~\ref{fig:residue2} (bottom panel) the histograms associated with $D_{\sigma}$ and $D_{\sigma}^\prime$. The median value and standard deviation of $D_{\sigma}$ are 46~\% and 42\,\%, respectively, while for $D_{\sigma}^\prime$ they are equal to  $-$12\,\% and 17\,\%, respectively.

As   mentioned for the scaling of $\taueff$ (see Sect.~\ref{comparision_taueff}), for both scaling relations, 
the dispersion and deviation with the measurements  can in part arise  from the fact that we observed an inhomogeneous sample of stars, in particular, stars with a different surface metal abundances. Indeed, the amplitude of the granulation background is expected to depend on the surface metal  abundance \citep[for a particular low-metal F-type star see][]{Ludwig09}. Furthermore, for the new scaling relation an  rms error of 100~K in $\teff$ and a rms error 0.1~dex on $\log g$ results for $z_3$ in a typical error  about 12~\% for RG stars and about $10$~\% for a typical MS (these typical relative dispersions are shown in Fig.~\ref{fig:sigma_z_Mascaling}). On the other hand, the  uncertainties associated with $\teff$ have no direct impact on the classical scaling relation given by \eq{c_3} since the term $\teff^{3/2} M^{-1/2}$ is estimated using only seismic constraints (see \eq{scaling_teff_M}).



Compared with the new scaling relation, the classical one results in a smaller difference with the observations.  However,  the deviations of the two scaling relations  from the measurements  are found to depend on $\teff$. The highest deviations are obtained for the F-dwarf stars ($\teff$=6\,000~K - 7\,500~K, see Fig.~\ref{diff:temp} and Sect.~\ref{discussion}). As discussed in Sect.~\ref{discussion}, this is very  likely a consequence of the lack of modelling of the  impact of magnetic activity on the granulation background. 


 As stressed in \PI, our theoretical calculations are expected to be valid for  stars with a low level of activity. If we exclude  the F-dwarf stars from our sample, the median deviation of the new scaling relation w.r.t the measurements is $-$2\,\% ($\pm$~30\,\%), while for the classical scaling relation it is equal to $-$19\,\% ($\pm$~18\,\%).
In that case, the new scaling relation results in a lower deviation. 
However, the difference between the median value of $D_{\sigma}$ and this of $D_{\sigma}^\prime$  is smaller than   the  standard deviation of  $D_{\sigma}$. Therefore, it is not possible to  distinguish the new theoretical scaling relation from the classical one.  


 
 In conclusion,  as the classical scaling relation, our theoretical scaling relation is compatible with the observations, but we cannot confirm the dependence on $\Ma$. 
 Observations of K-dwarf stars ($\teff=3\,500 - 5\,000$~K) could in principle help to check the dependence of the theoretical scaling relation on $\Ma$. Indeed, for instance the 3D model with $\teff \simeq$~4\,500~K and $\log g$=4.0 (K dwarf) has $\numax=$1.3~mHz $\Ma \simeq $ 0.18,  and $\sigma \simeq$18~ppm, while the 3D model $\teff \simeq$~5~900~K and same $\log g$ (G dwarf) has $\numax=$1.1~mHz $\Ma \simeq $ 0.31,  and $\sigma \simeq $110~ppm. The relative difference in $\sigma$ between the K dwarf model and the G-dwarf model is 84\,\%. This is much higher than the dispersion in $D_{\sigma}$ and $D_{\sigma}^\prime$.

\section{Removing the degeneracy with the mass and the radius}
\label{degeneracy}


As seen in Sect~\ref{comparision_individual}, the individual theoretical values of $\sigma$ are found to scale as $z_3^p$ with the slope $p$=1.10. 
As we will show now, the deviation of the individual values of $\sigma$ from a linear scaling with $z_3$  is for a large part due to the considerable degeneracy that occurs for red giants between  $M$ and $R$. 
Indeed, the theoretical values of $\sigma$ scale as ${\cal N}_g^{-1/2}$, and hence as the stellar radius $R_s$ (see Eqs.~(\ref{sigma_tau}) and (\ref{N_g})). Furthermore,  $z_3$ scales as  $M^{-1/2}$. Therefore theoretical values of $\sigma$ and $z_3$ directly depend on the masses and radii attributed to the 3D models. 
However, two red giants with same $\teff$ and  $\log~g$ can have very different values of $R$ and $M$. 
 Furthermore, the masses and radii attributed to our 3D models were obtained from  a grid of standard stellar models with fixed physical assumptions, and all of these models are in the pre-helium-burning phase, which is not  the case for all observed RG stars.

 When we multiply theoretical $\sigma$ by $R_s/R_\odot$, we obtain a quantity that does no longer depend on the radius attributed to the 3D model. Furthermore,  the quantity  $z_4 \equiv z_3 \, (R_s/R_\odot)$  scales as $g^{-1/2}$. As a consequence, $z_4$  does not depend on the mass attributed to the 3D model.
To remove possible bias introduced by the determination of the masses and radii of the 3D models we must therefore compare theoretical values of  ${\tilde \sigma}$ as a function of $z_4$ with the measurements multiplied by the star radii. To do this, we  need to determine the radii of the observed targets. Combining the scaling relation for $\numax$ with the one for $\Delta \nu$ gives \citep[see e.g.][]{Stello09,Kallinger10,Mosser10}
\eqn{
{R_s \over  R_\odot } =   \left ( { \numax \over \numaxref }  \right ) \,  \left ( { \Delta \nu \over \deltanuref  }  \right )^{-2} \, \left ( \teff  \over  \teffsun \right )^{1/2} \; .
\label{eq:radius_scaling}
}
Multiplying \eq{sigma_scaling_2} by \eq{eq:radius_scaling} gives   the  scaling  relation for ${\tilde \sigma} =  (R_s/R_\odot) \, \sigma$   with the help of \eq{scaling_teff_M}
\eqn{
{\tilde \sigma} \propto z_4 =  \left ( { \numaxref \over  \numax  }  \right ) \,  \left ( { {f(\Ma)} \over {f(\MaO) } } \right )^2 \; .
\label{sigma_scaling_3}
}
To compare theoretical ${\tilde \sigma}$ with the measurements, we multiply the measured $\sigma$ by the ratio $R_s/R_\odot$ given by \eq{eq:radius_scaling}. 
We have plotted  theoretical and measured values of ${\tilde \sigma}$ in Fig.~\ref{fig:sigma_z_2}.
The individual theoretical values of ${\tilde \sigma}$ are found to scale as   $z_4^p$ with $p = 1.03$ and are therefore  better  aligned with the measurements than those of $\sigma$.

\fig{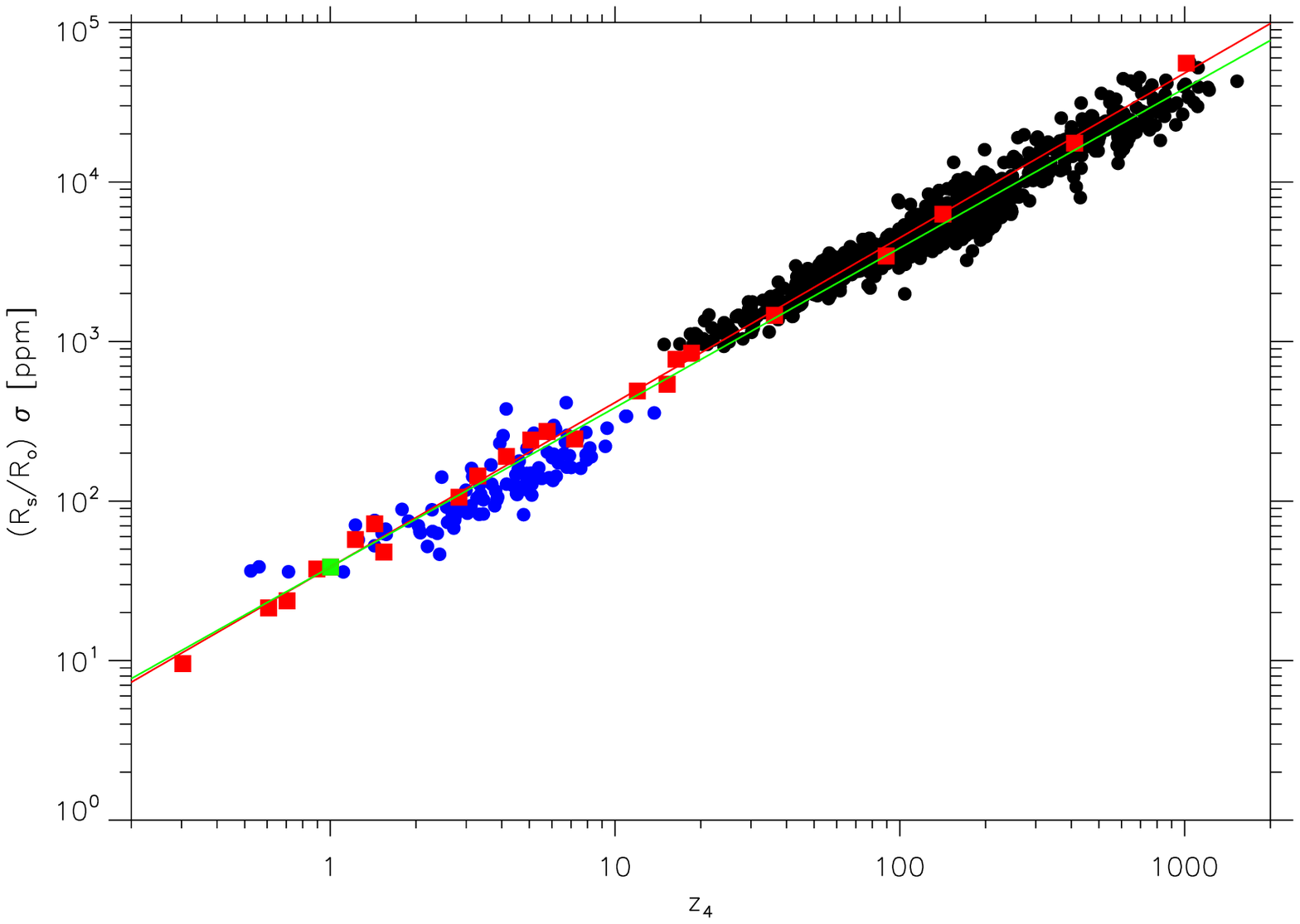}{ ${\tilde \sigma} \equiv  (R_s/R_\odot) \, \sigma $ as a function the quantity $z_4$ given by \eq{sigma_scaling_3}. The symbols have the same meaning as in Fig.~\ref{HR}. The green line corresponds to a linear scaling with $z_4$ and the red one to a power law of the form  $ z_4^{p}$ where the slope $p = 1.03$ is obtained by fitting the individual theoretical values of ${\tilde \sigma}$ (red squares). } {fig:sigma_z_2}

\end{document}